\def\spose#1{\hbox to 0pt{#1\hss}}

\def\multleft#1{\hbox to size{\vbox {\halign {\lft{##}\cr #1}}\hfill}\par}
\def\multright#1{\hbox to size{\vbox {\halign {\rt{##}\cr #1}}\hfill}\par}

\def\degmark{^\circ}
\def\boxit#1{\vbox{\hrule\hbox{\vrule\kern3pt\vbox{\kern3pt
          #1 \kern3pt}\kern3pt\vrule}\hrule}}

\def\cm{{\rm\thinspace cm}}

\def\erg{{\rm\thinspace erg}}

\def\K{{\rm\thinspace K}}

\def\km{{\rm\thinspace km}}
\def\kpc{{\rm\thinspace kpc}}

\def\s{{\rm\thinspace s}}
\def\yr{{\rm\thinspace yr}}

\def\pcmcu{\hbox{$\cm^{-3}\,$}}

\def\ergps{\hbox{$\erg\s^{-1}\,$}}

\def\kmps{\hbox{$\km\s^{-1}\,$}}

\def\C00{Churazov et al. (2000)}
\def\K00{Br\"uggen \& Kaiser (2000)}

\documentstyle[graphicx]{mn}

\title{The hydrodynamics of dead radio galaxies}

\author[Reynolds, Heinz \& Begelman]{Christopher~S.~Reynolds$^{1,2}$
\thanks{Hubble Fellow while at University of Colorado}, Sebastian
Heinz$^3$, and Mitchell C. Begelman$^{1,4}$\\ 
{$^1$JILA, Campus Box 440, University of Colorado, Boulder CO~80309, USA}\\
{$^2$Dept. of Astronomy, University of Maryland, College Park, MD~20742, USA}\\
{$^3$Max-Planck-Institut f\"{u}r Astrophysik,
Karl-Schwarzschild-Str. 1, 85740 Garching, Germany}\\ 
{$^4$Dept. of
Astrophysical and Planetary Sciences, University of Colorado, Boulder
CO~80309, USA}\\ }

\begin{document}

\maketitle

\begin{abstract}
We present a numerical investigation of dead, or relic, radio galaxies
and the environmental impact that radio galaxy activity has on the
host galaxy or galaxy cluster.  We perform axisymmetric hydrodynamical
calculations of light, supersonic, back-to-back jets propagating in a
$\beta$-model galaxy/cluster atmosphere.  We then shut down the jet
activity and let the resulting structure evolve passively.  The dead
source undergoes an initial phase of pressure driven expansion until
it achieves pressure equilibrium with its surroundings.  Thereafter,
buoyancy forces drive the evolution and lead to the formation of two
oppositely directed plumes that float high into the galaxy/cluster
atmosphere.  These plumes entrain a significant amount of low entropy
material from the galaxy/cluster core and lift it high into the
atmosphere.  An important result is that a large fraction (at least
half) of the energy injected by the jet activity is thermalized in the
ISM/ICM core.  The whole ISM/ICM atmosphere inflates in order to
regain hydrostatic equilibrium.  This inflation is mediated by an
approximately spherical disturbance which propagates into the
atmosphere at the sound speed.  The fact that such a large fraction of
the injected energy is thermalized suggests that radio galaxies may
have an important role in the overall energy budget of rich ISM/ICM
atmospheres.  In particular, they may act as a strong and highly
time-dependent source of negative feedback for galaxy/cluster cooling
flows.
\end{abstract}


\section{Introduction}

There is now overwhelming observational and theoretical evidence that
powerful radio galaxies possess highly collimated and relativistic
twin jets of matter that emerge from a central active galactic nucleus
(AGN).  In the powerful Faranoff-Riley type II radio galaxies (FR-II;
Fanaroff \& Riley 1974), the jets are thought to remain relativistic
along most of their length before terminating in a series of shocks
resulting from interaction with the surrounding material.  A substantial
amount of theoretical and numerical (hydrodynamic) work suggests that,
after passing through the terminal shock, the spent jet material
inflates a broad cocoon which encases the jets (e.g., see Scheuer 1974;
Norman et al. 1982; Lind et al. 1989; Begelman \& Cioffi 1989; Cioffi \&
Blondin 1992; Hooda, Mangalam \& Wiita 1994).

In the early life of a powerful source, the cocoon is probably
overpressured (Begelman \& Cioffi 1989) with respect to the surrounding
ambient material (either the interstellar medium [ISM] of the host
galaxy, or the intracluster medium [ICM] of the host cluster).  The
cocoon then undergoes supersonic pressure-driven expansion into the
ambient medium, sweeping the ambient medium into a shocked shell.  A
contact discontinuity separates the relativistic cocoon material from
the shocked ambient material.  From the onset of activity,
Kelvin-Helmholtz (KH) instabilities act to shred the contact
discontinuity.  If the ambient medium has a density profile
characterized by $\rho\propto r^{-\alpha}$ where $\alpha>2$, the contact
discontinuity will accelerate and hence will be Rayleigh-Taylor (RT)
unstable.  However, in the more physical case of $\alpha<2$ the contact
discontinuity will possess an initial deceleration, which is sufficient
to stabilize the contact discontinuity against RT modes.  Only at later
times, once the cocoon comes into approximate pressure balance with its
surroundings, will the deceleration of the contact discontinuity no
longer exceed the local gravitational acceleration, thereby rendering it
RT unstable.  Eventually, AGN activity will cease and the jets will turn
off.  Although this stage of a radio galaxy's life has been little
studied, it seems likely that the combined action of the KH and RT
instabilities will transform the remnant of the cocoon into `plumes'
that rise in the potential of the galaxy/cluster under the action of
buoyancy forces (Gull \& Northover 1973; Churazov et al. 2000; Br\"uggen
\& Kaiser 2000).

Observationally, radio galaxies are often seen to be associated with
galaxy or cluster cooling flows.  This raises an obvious and interesting
set of questions.  To what extent is radio galaxy activity a natural
{\it response} to the cooling flow phenomenon?  Do radio galaxies act
back on their environment to partially offset the cooling flow, i.e.,
are radio galaxies a dramatic manifestation of le Chatelier's
principle?  Is radio galaxy activity a crucial component for our
understanding of galaxy clusters?  Since it takes only a small fraction
of the cooling flow mass to fuel a powerful AGN, it is easy to see how
radio galaxy activity can result from a cooling flow.  However, the
impact of a radio galaxy on its environment is much less well
understood.

In this paper, we use hydrodynamic simulations to study the evolution of
a radio galaxy situated at the center of a galaxy/cluster.  In
particular, we follow the evolution of the source at times after the
jets have terminated.  We examine the interaction of the radio galaxy
with the ISM/ICM and assess the environmental impact of the AGN
activity.  Section~2 presents our set of hydrodynamic simulations and
focuses on some numerical issues.  In Section~3, we briefly discuss the
`active' phase of the source.  Section~4 presents our results for dead
sources (i.e., the passive phase after the jets have turned off), and
Section~5 discusses some astrophysical implications as well as the
limitations of our calculations.  Section~6 draws our conclusions
together.

\section{The hydrodynamic simulations}

\subsection{Basics and simulation setup}

We model a source with back-to-back supersonic jets situated at the center
of a spherical gaseous atmosphere.  To permit parameter studies of various
high resolution simulations, we make the assumption of axisymmetry and
neglect magnetic fields.  We discuss these assumptions in Section~5.  The
simulations were performed using the ZEUS-3D code (Stone \& Norman
1992a,1992b; Clarke, Norman \& Fiedler 1994) in its 2D (axisymmetric) pure
hydrodynamic mode.  ZEUS-3D is an explicit Eulerian finite difference code,
formally of second order, which uses an artificial viscosity to reproduce
shocks.  Major advantages of this code are its flexibility in the choice of
the computational grid, and the fact that it has been extensively tested.
The simulations were performed on a Sun Microsystems Ultra 60 workstation.

Unlike many previous simulations, we model both jets and make no assumption
regarding reflection symmetry in the plane normal to the jets.  This allows
us to study phenomena associated with the collision of the two backflows.
Our simulations were performed in spherical polar coordinates $(r, \theta,
\phi)$, and the computational domain was the region $r\in (0.1,10)$, except
for one run (Run~4 below) which had the domain $r\in (0.1,30)$.  All gas
within the calculation was assumed to have an adiabatic index of
$\gamma=5/3$.  The ambient galaxy/cluster gas was assumed to be
initially in an isothermal configuration with a (adiabatic) sound speed
$c_{\rm s}=1$ and a density profile given by a $\beta$-model with
$\beta=0.5$:
\begin{equation}
\rho(r)=\frac{1}{[1+(r/r_0)^2]^{3/4}},
\end{equation}
where the core radius was set to $r_0=2$.  The gravitational potential,
$\Phi$, was set so as to keep this ambient material in hydrostatic
equilibrium
\begin{equation}
\frac{{\rm d}\Phi}{{\rm d}r}=\frac{1}{\rho}\frac{{\rm d}p}{{\rm d}r},
\end{equation}
which gives (to within an additive constant)
\begin{equation}
\Phi=\frac{c_{\rm s}^2}{\gamma}\ln(\rho).
\end{equation}
It is assumed that the gravitational potential is dominated by a
stationary distribution of background dark matter, i.e., the
self-gravity of the gas is negligible.  This reasonable assumption
prevents us from having to solve Poisson's equation as part of the
hydrodynamic simulations.

We inject back-to-back jets into this background medium via the use of
inflow boundary conditions at $r=0.1$.  The jets are initially conical
with a $15^\circ$ half opening angle, and in pressure equilibrium with
the ambient medium.  They are given an initial density of $\rho_{\rm
jet}=0.01$ and Mach number (with respect to the sound speed of the
injected jet material) of ${\cal M}=10$.  As discussed in the Appendix,
these parameters are chosen so that the Kelvin-Helmzholtz growth rate at
the cocoon-ICM contact discontinuity in this {\it non-relativistic}
simulation is approximately the same as the corresponding growth rate in
the real {\it relativistic} situation.  At time $t=1$, the jet activity
is stopped and the inner boundary is made reflecting.  The source is
then allowed to evolve passively.

We can relate quantities within the code to physical quantities by
fixing the parameters of the background medium.  Suppose we set
$r_0=100\kpc$, $c_{\rm ISM}=1000\kmps$ and a central number density of
$n_0=0.01\pcmcu$ --- values representative of a rich galaxy cluster.
Then, one code unit of time corresponds to 50\,Myr.  The total kinetic
luminosity of the source is then $9.3\times 10^{45}\ergps$.  These
parameters are relevant for powerful sources in rich galaxy clusters
such as Cygnus~A (Arnaud et al. 1984; Reynolds \& Fabian 1996) and
3C~295.

Unless otherwise stated, this paper shall use the above scalings when
converting our simulation results into quantities that can be compared
directly with real systems.  However, in the present universe at least,
many radio galaxies are found in smaller galaxy clusters or poor groups.
Taking objects like Hydra~A and Virgo~A as prototypes, the typical
relevant ISM/ICM parameters are $r_0=10\kpc$, $c_{\rm ISM}=500\kmps$ and
$n_0=0.1\pcmcu$ (e.g., see the detailed analysis of the {\it Chandra
X-ray Observatory} data for Hydra~A by David et al. 2000).  In this case,
one time unit corresponds to 10\,Myr and the total kinetic luminosity of
the source is $1.16\times 10^{44}\ergps$.  Of course, it must be
appreciated that these $\beta$-models are very crude parameterizations
of the real ISM/ICM density structure.

\subsection{The need for adequate numerical resolution}

During the active phase, the large scale evolution and properties of the
cocoon are strongly influenced by much smaller scale shock structures that
appear in the jet.  As will be discussed in more detail below, a conical
shock appears in the jet which sprays the jet thrust over a large working
surface (which expands almost self-similarly).  Failure to resolve this
shock structure in a simulation leads to the jet thrust being deposited
over a very small area which leads to the jet rapidly `drilling' a narrow
cavity through the ICM.

Initially, we performed simulations with $r_{\rm out}=10$ at three
different resolutions: $n_{\rm r}\times n_{\theta}=300\times 300$
(Run~1), $600\times 600$ (Run~2), and $1200\times 1200$ (Run~3).  The
resolution was slightly enhanced at small $r$ and along the jet axes.
We find that Run~1 does indeed fail to resolve the jet-shock
structure.  On the other hand, Run~2 and Run~3 both well resolve the
internal shocks within the jet and produce cocoons that expand in
almost a self-similar manner, at least during the early active phase.
The similarity of these two simulations gives us some confidence that
we have achieved a degree of numerical convergence.

We note one additional numerical technicality.  Our simulations possess
a relatively high dynamical range in $r$.  However, to maintain adequate
resolution on large scales, we choose not to use a logarithmic
(``scaled'') grid for the $r$ variable.  Instead, we space our grid
using a more general geometric progression (i.e. ``ratioed'' scaling).
While allowing us to maintain adequate resolution on the scales that
matter to our problem, it has the undesirable effect of forcing highly
elongated grid elements to exist in the very central regions of the
simulation.  One might then worry about the effects of an effective
anisotropic artificial viscosity in this inner region.  However, this is
acceptable since we are not interested primarily in the detailed
hydrodynamic properties of these inner regions.  One can view the
innermost region of our grid as a buffer zone, which connects smoothly
to the interesting region of the grid, while protecting it from the hard
inner boundary.  Experiments with high resolution simulations using a
logarithmic $r$-grid give us confidence that this is not a problem.

For the rest of this paper, we discuss results from Run~2 (i.e., $r_{\rm
out}=10$ with a grid size of $n_{\rm r}\times n_{\theta}=600\times
600$), and one other simulation.  This final simulation (Run~4) has a
larger outer radius ($r_{\rm out}=30$) and a larger computational grid
($n_{\rm r}\times n_{\theta}=1200\times 600$).   In the spatial region
covered by both Run~2 and Run~4, the resolutions are approximately (but
not exactly) the same.

\section{The active phase}

\begin{figure*}
\hbox{
\includegraphics[width=0.45\textwidth]{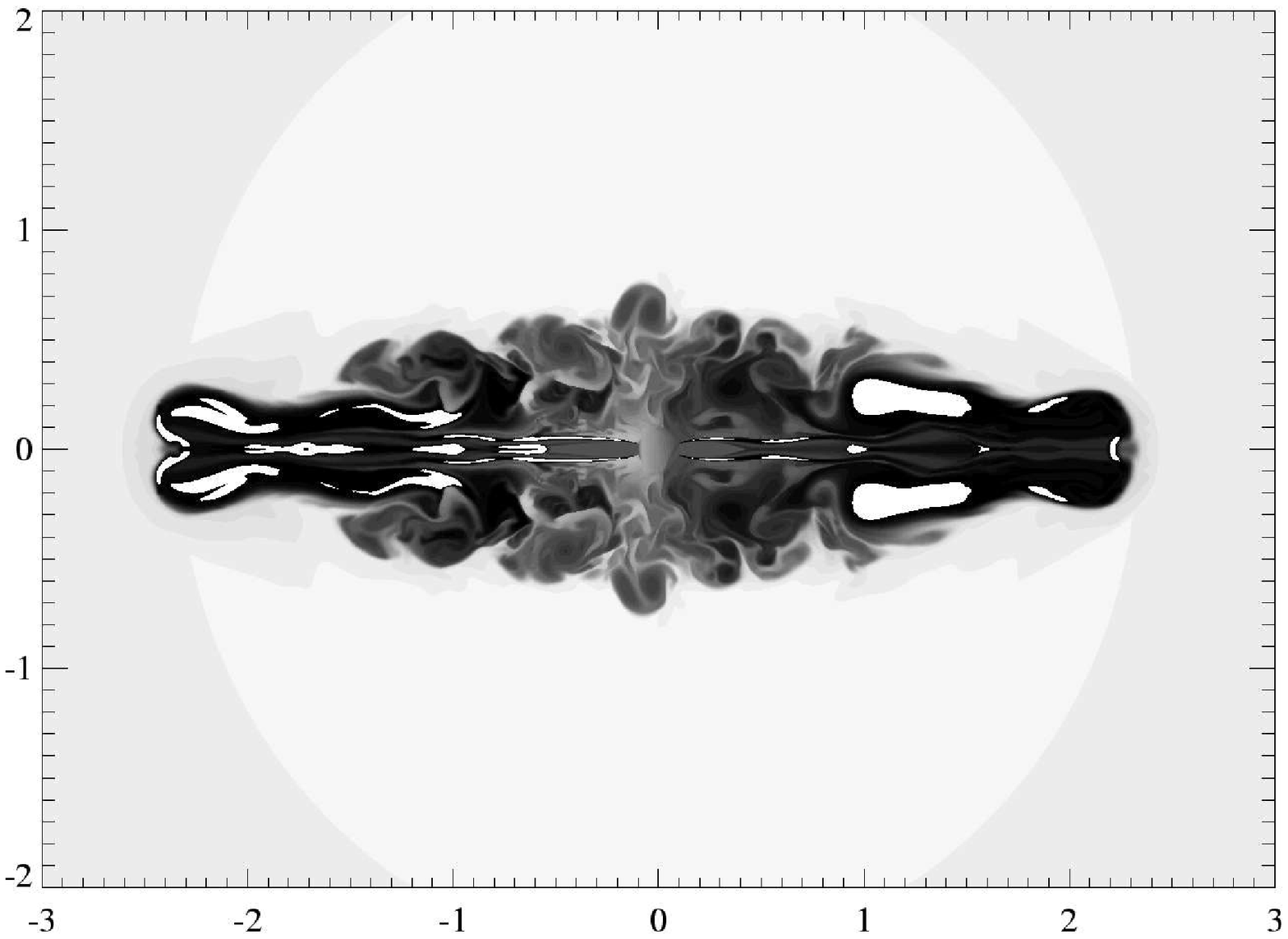}
\includegraphics[width=0.45\textwidth]{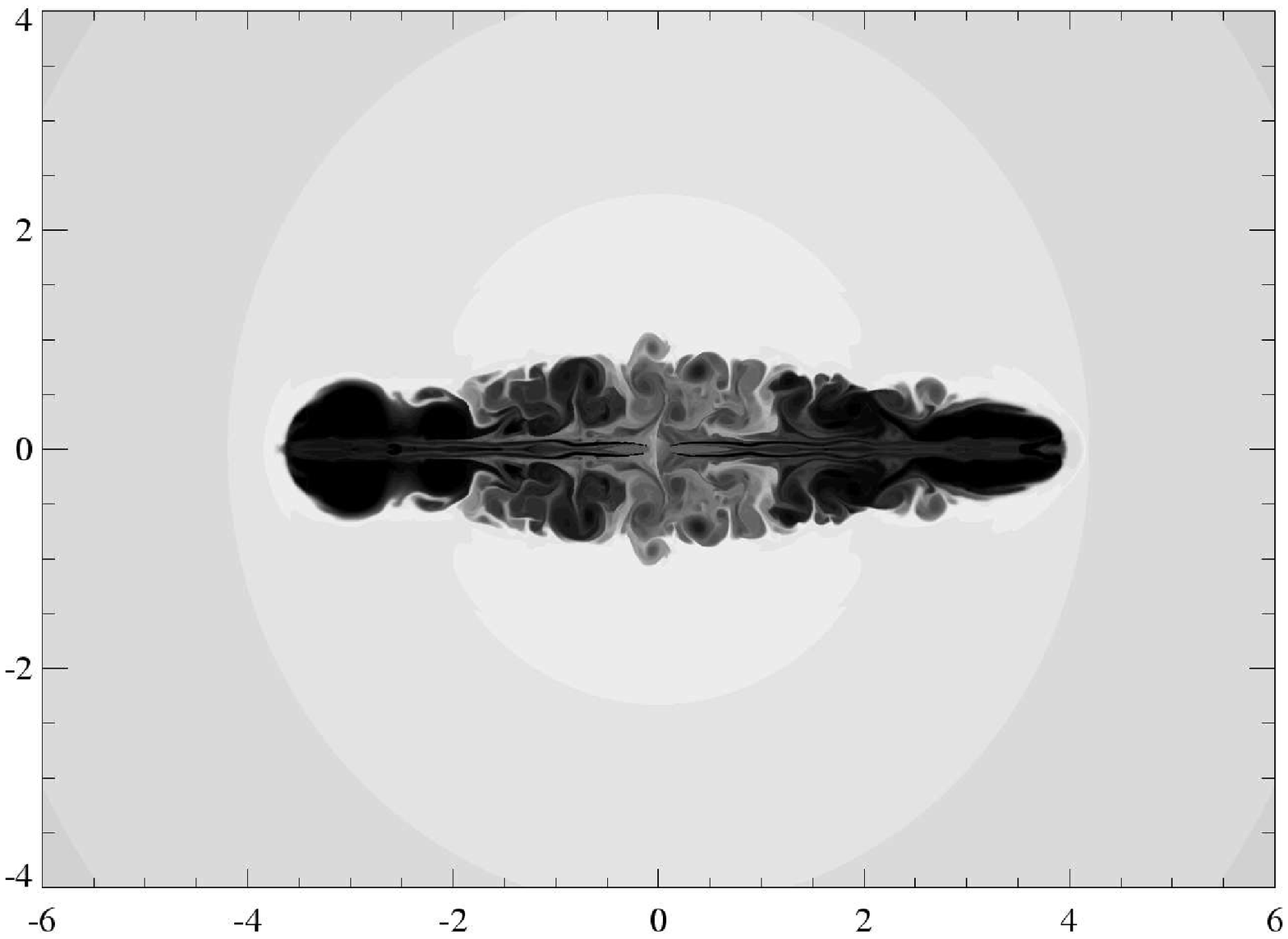}
}
\hbox{
\includegraphics[width=0.45\textwidth]{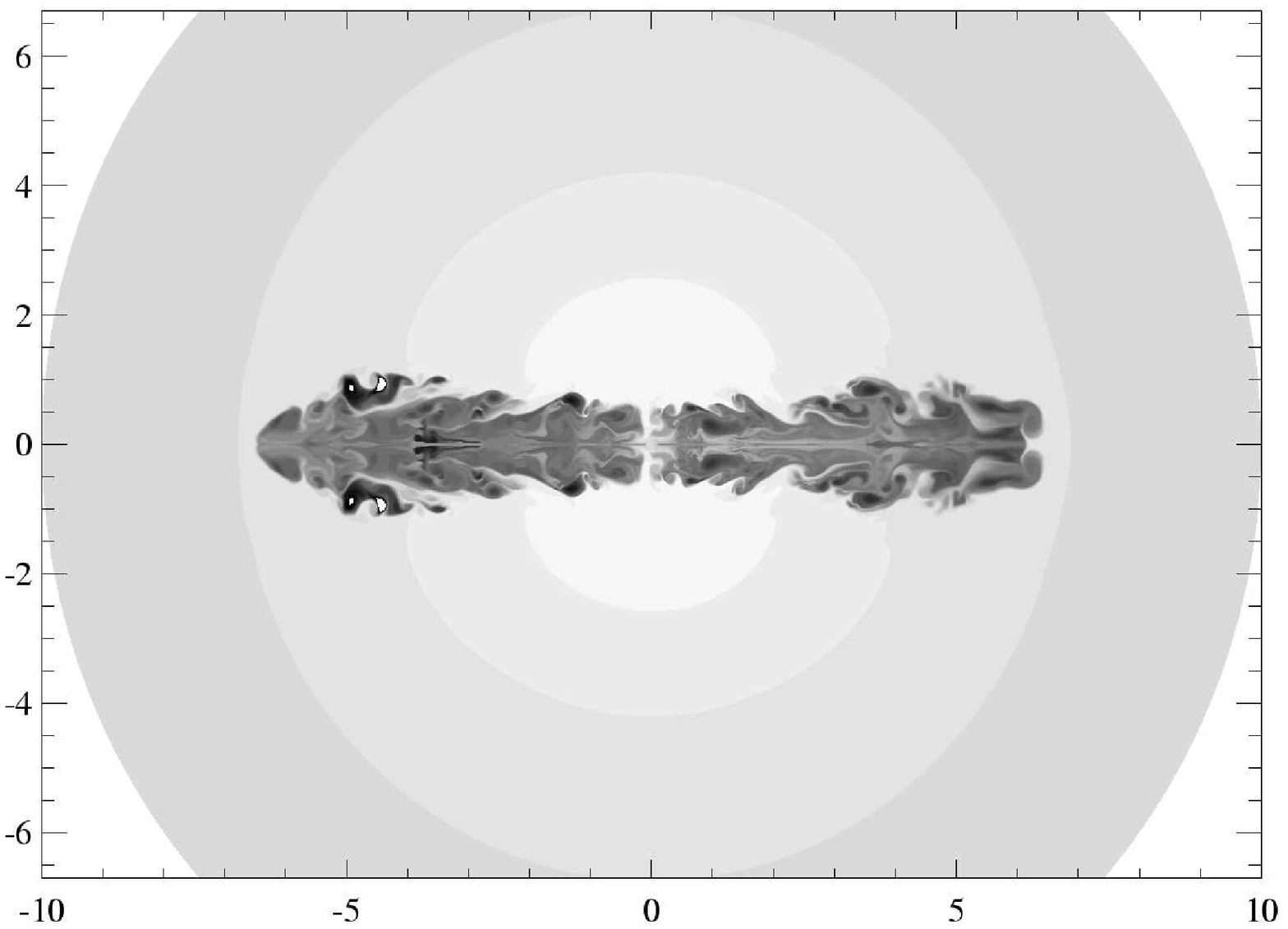}
\includegraphics[width=0.45\textwidth]{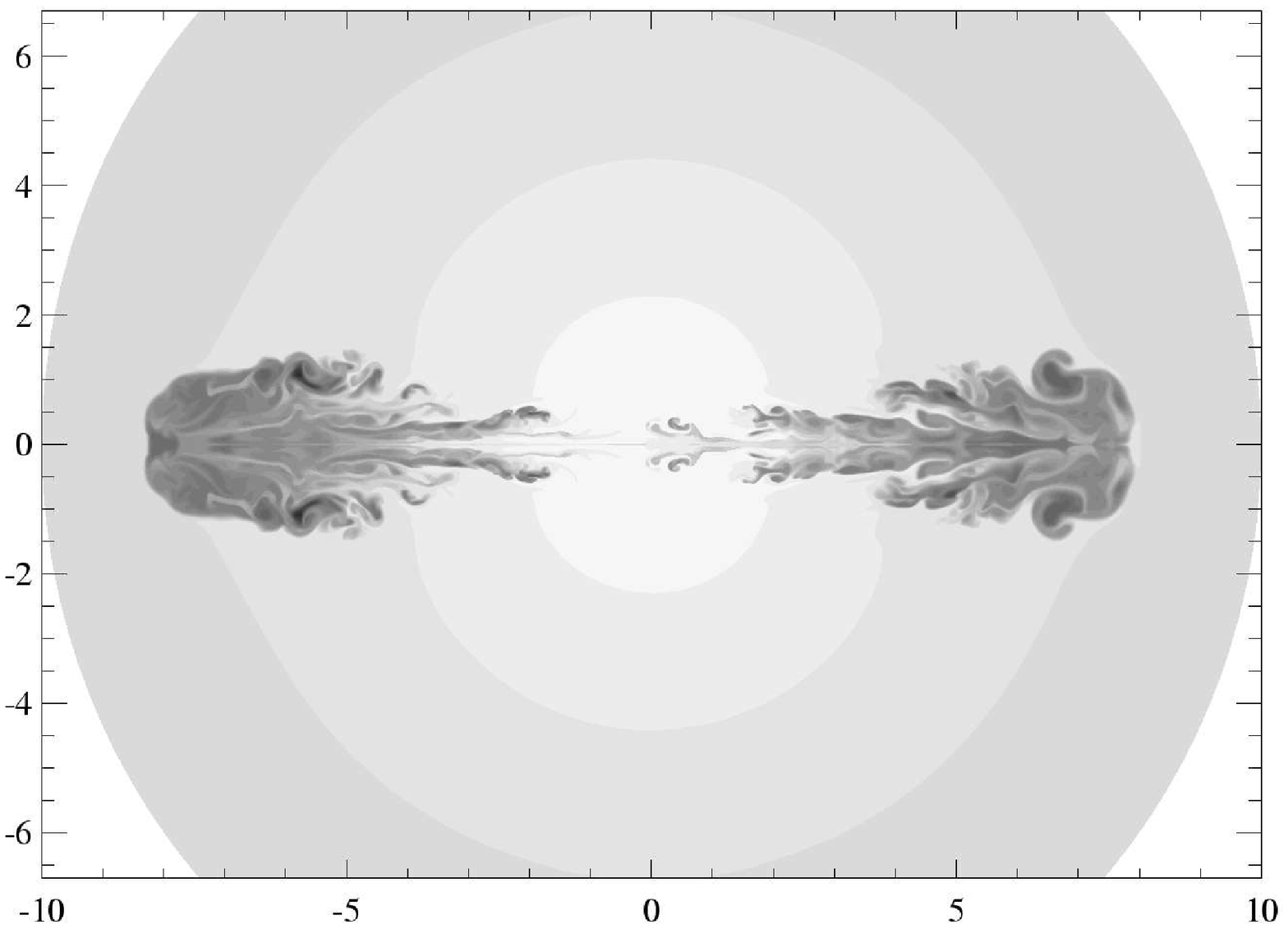}
}
\caption{Specific entropy maps for $t=0.50$, $t=1.0$, $t=3.0$ and
$t=5.0$ for Run~2.  As the same greyscale levels are used for all four
panels, one can clearly see the decrease in the specific entropy of the
cocoon it rises and entrains ambient thermal plasma.  Note the change of
scale in the first three of these plots.\label{fig:entropy_map_1}}
\end{figure*}

\begin{figure*}
\hbox{
\includegraphics[width=0.45\textwidth]{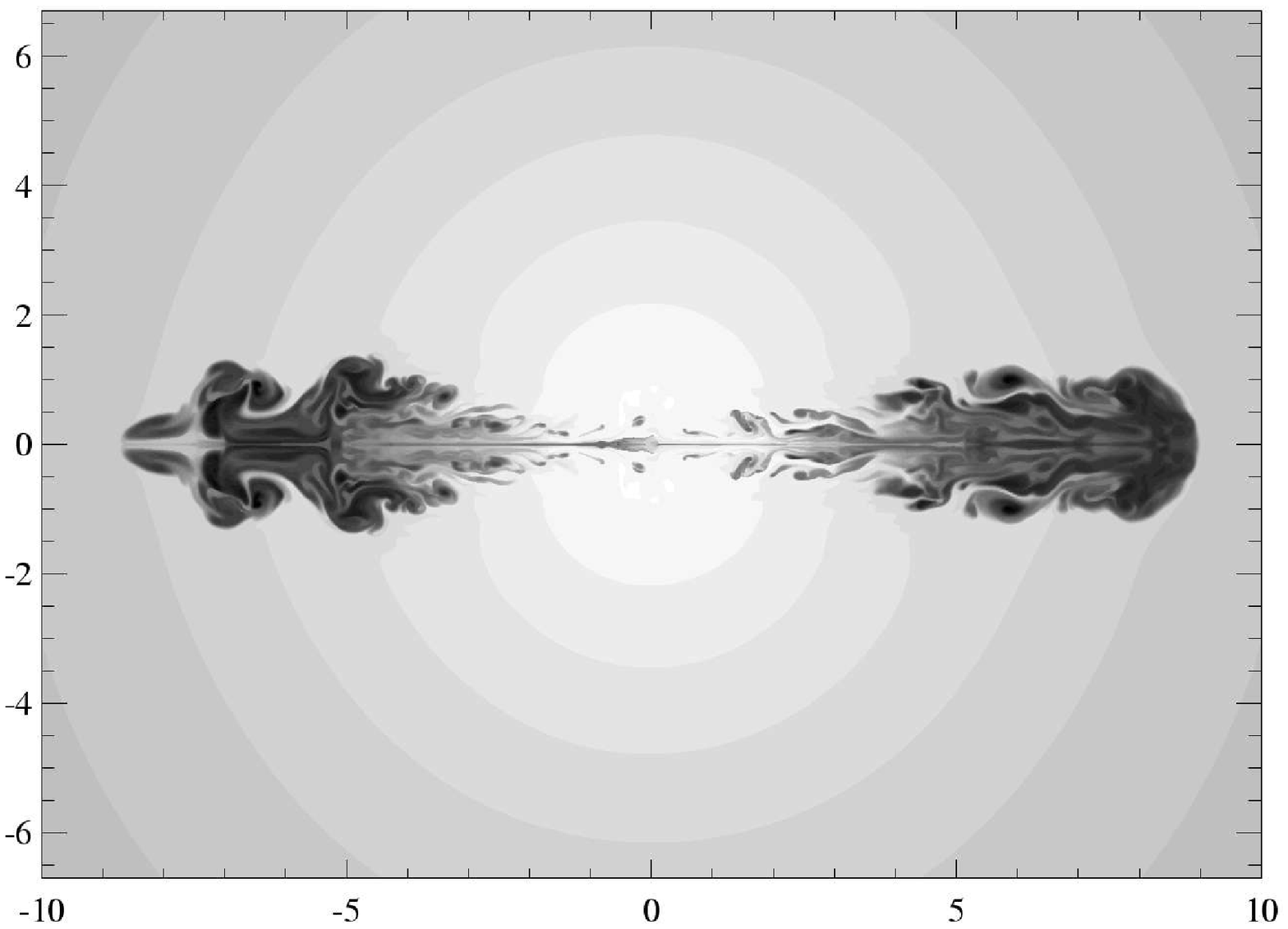}
\includegraphics[width=0.45\textwidth]{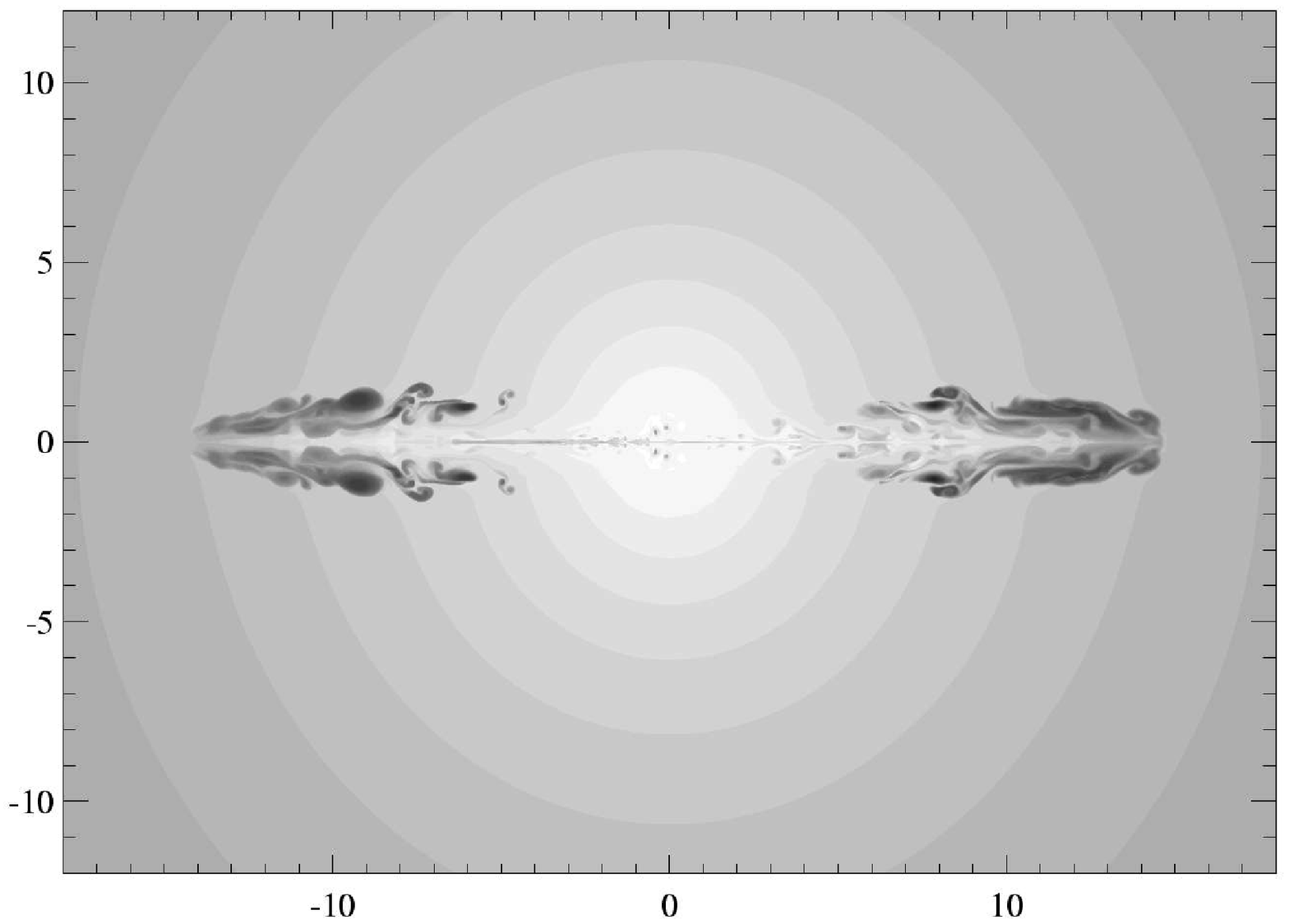}
}
\hbox{
\includegraphics[width=0.45\textwidth]{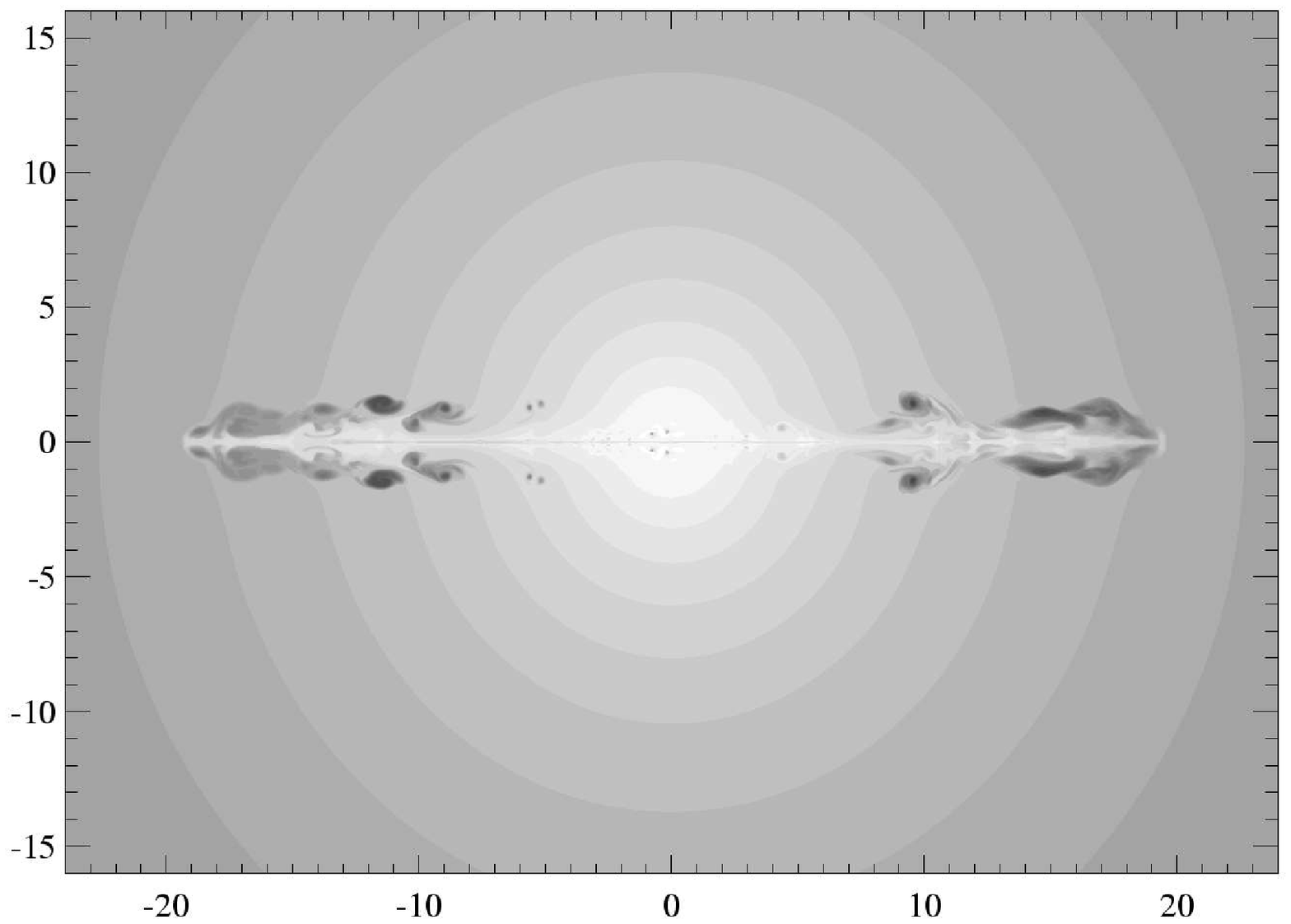}
\includegraphics[width=0.45\textwidth]{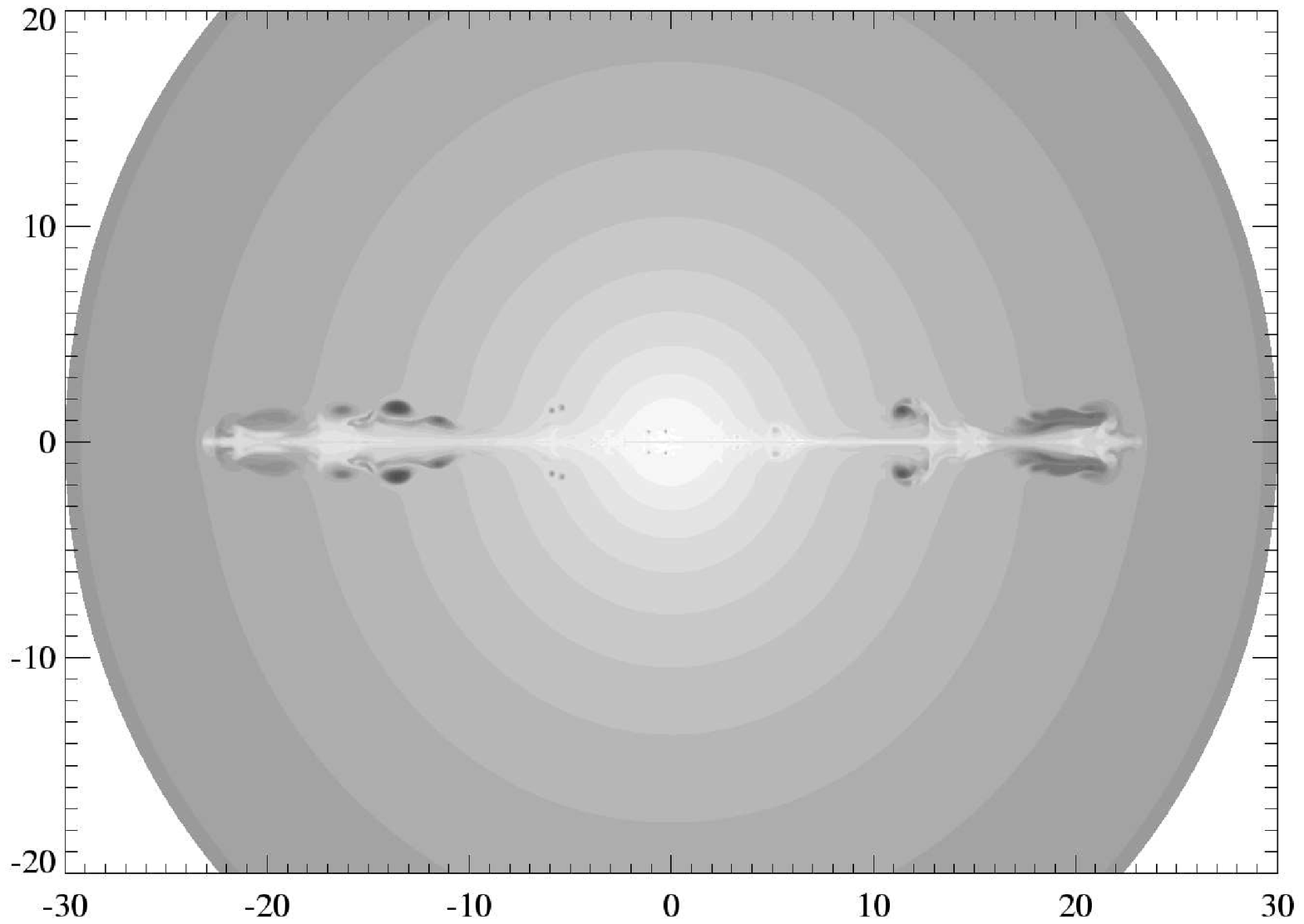}
}
\caption{Specific entropy maps for $t=5.0$, $t=10.0$, $t=15.0$ and
$t=20.0$ for Run~4.  It can be seen that low entropy
material, from the center of the galaxy/cluster atmosphere, is being
lifted by the buoyant plumes.  Note the change in scale between all of
the four panels.\label{fig:entropy_map_2}}
\end{figure*}

There is already a substantial body of numerical work on the
hydrodynamics, and magnetohydrodynamics, of active radio galaxy jets.
Thus, we shall discuss only briefly our simulation results for the
active phase of our source.  In a companion paper (Reynolds, Heinz \&
Begelman 2000) we focus on the X-ray appearance of such active sources.

Our results during the active phase are very similar to those obtained by
Lind et al. (1989), who performed axisymmetric hydrodynamic simulations of
a single jet propagating through a uniform medium.  The initially conical
jet is focused by the external pressure and suffers a series of oblique
shocks.  Near the working surface with the background medium, a strong
annular shock deflects most of the flow into a wide fan.  Thus, the jet
thrust is deposited over a large area, thereby reducing the advance speed
of the jet below the value that would be obtained if the jet just `drilled'
into the background medium.  This is the 2-dimensional equivalent of the
so-called ``dentist drill effect'' (Scheuer 1974).  After passing through
the annular shock, the spent jet material enters a mildly supersonic
backflow which gradually decelerates through a series of weak shocks.
Powerful Kelvin-Helmholtz (KH) instabilities exist along the contact
discontinuity between the backflowing jet material and the
shocked/compressed ambient material.  These KH instabilities, which become
more powerful as the backflow decelerates toward the region where the two
backflows collide, act to mix ambient material into the cocoon.  These
instabilities can be seen clearly in Fig.~\ref{fig:entropy_map_1}, which
shows specific entropy maps ($s=p/\rho^{5/3}$; note that we neglect
logarithms and constants in our definition of the specific entropy) for
Run~2 at times $t=0.5$, $t=1.0$, $t=3.0$ and $t=5.0$.  During the active
phase, the structure expands approximately self-similarly, at least as long
as the cocoon is still sufficiently overpressured as to undergo supersonic
lateral expansion into the ambient material.

It is worth noting explicitly that the backflow in the cocoon has a
speed comparable to the sound speed of the cocoon material.  In code
units, the advance speed of the working surface of the jet is $\sim 3$
(i.e., a Mach number of 3 with respect to the sound speed of the
undisturbed ambient material).  This should be compared with a backflow
speed of $\sim 20-30$ which exists in a large portion of the cocoon.
Translated into terms applicable to real radio galaxies, the backflow
speed would be a significant fraction of the speed of light (since the
spent jet material is a relativistic plasma with a sound speed of
$c/\sqrt{3}$).

\section{The passive phase}

As discussed in Section~2, we turn off the jets after $t=1$ and allow
the existing structure to evolve.  Both Run~2 and Run~4 display very
similar behavior during the passive phase although, of course, Run~4 can
be followed to much later times due to the larger spatial domain of the
calculation (see Fig.~2).  We will describe our findings based on
Run~4. It should be noted that \K00 and \C00 have also simulated the
late evolution of radio galaxies. Their simulations make the assumption
that the source evolution before the jets turn off is governed by simple
supersonic expansion of the lobes and consequently, their initial
conditions for the passive phase are based on relatively simple
geometries; namely, spherical and ellipsoidal bubbles placed in
isothermal atmospheres. However, turbulence and powerful circulatory
motion seems to be present at all times inside the active cocoon, which
leads to significant excitation of the Kelvin-Helmholtz instability
before the source turns off.  Whether, and by how much, the cocoon is
overpressured at the moment that the jet activity ceases can be
estimated by the simple analytic models of Begelman \& Cioffi (1989).

Thus, we extend the work of \C00 and \K00
by simulating the active phase of the source and then shutting the jets
off, rather than modelling the late stages of evolution by letting a
static bubble evolve under the action of buoyancy.  We also simulate a
substantially larger portion of the cluster atmosphere while maintaining
the required high resolution, thereby allowing us to track the source
evolution at later times.  We consider these aspects to be important
extensions of these previous works.

\subsection{The stages of an inactive source's life}

As would be expected from simple qualitative arguments, turning off the
jets results in the almost immediate disappearance of internal shocks in
the jet channels, which subsequently collapse. The very rapid backflow
motions in the cocoon slow down accordingly, due to the lack of
high-inertia material and the disappearance of the terminal shock, which
previously re-directed the flow.  While the radial lobe expansion during
the active phase was governed mostly by the jet thrust, the absence of
ram pressure from the jets in the inactive phase results in a rapid
slowing of the cocoon head to subsonic speeds.  If the cocoon is still
significantly overpressured when the jet activity ceases, this subsonic
stage will be preceded by a Sedov expansion phase.

During the {\it early inactive phase} of our simulations, the source
is still slightly overpressured with respect to the ambient medium and
we do indeed find a Sedov phase, during which the cocoon expands in
order to achieve pressure balance.  When pressure balance is reached
(at $t\sim 1.5$), the driving force on the shocked/compressed ICM
shell ceases and this shell rapidly decelerates to subsonic speeds.
Once the lobes reach rough pressure equilibrium with their
surroundings, the dynamics of the lobes closely resemble a rising
bubble.  In the initial stages of this {\it late inactive, or buoyant,
stage}, the denser IGM begins to settle back into the center,
squirting the light cocoon material along the major axis of the
cocoon.  Eventually, the ICM core will reform and the cocoon will be
pinched off into two separate, buoyantly rising plumes.

\begin{figure}
\includegraphics[width=\columnwidth]{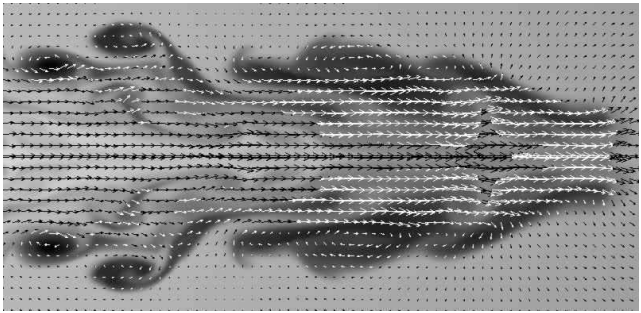}
\caption{Entropy map of a buoyant plume. The arrows correspond to
 the 2D velocity field.  \label{fig:plumemap}}
\end{figure}

Material that had been entrained previously is now carried along by each
plume, partly due to its own inertia and partly due to the significant
circulation inside the plume. This vorticity continues to drag ambient
material into the bubble at the plume's back end, where strong eddies
produce a jelly-fish-like structure (Fig.~\ref{fig:plumemap}). At the
head of the plume, a dense cap of swept up material forms, which is
pushed ahead.  

\begin{figure*}
\hbox{
\includegraphics[width=0.5\textwidth]{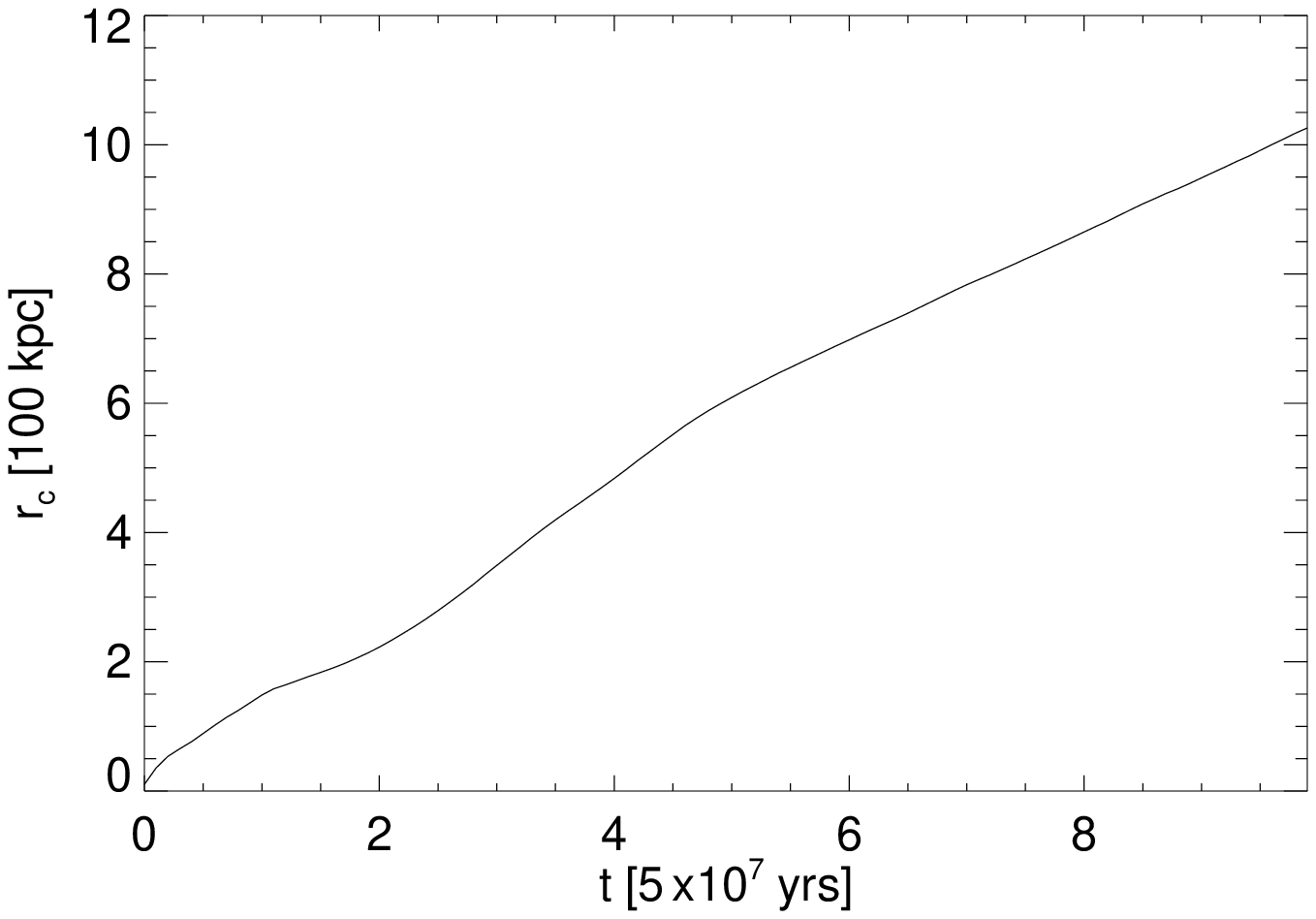}
\includegraphics[width=0.5\textwidth]{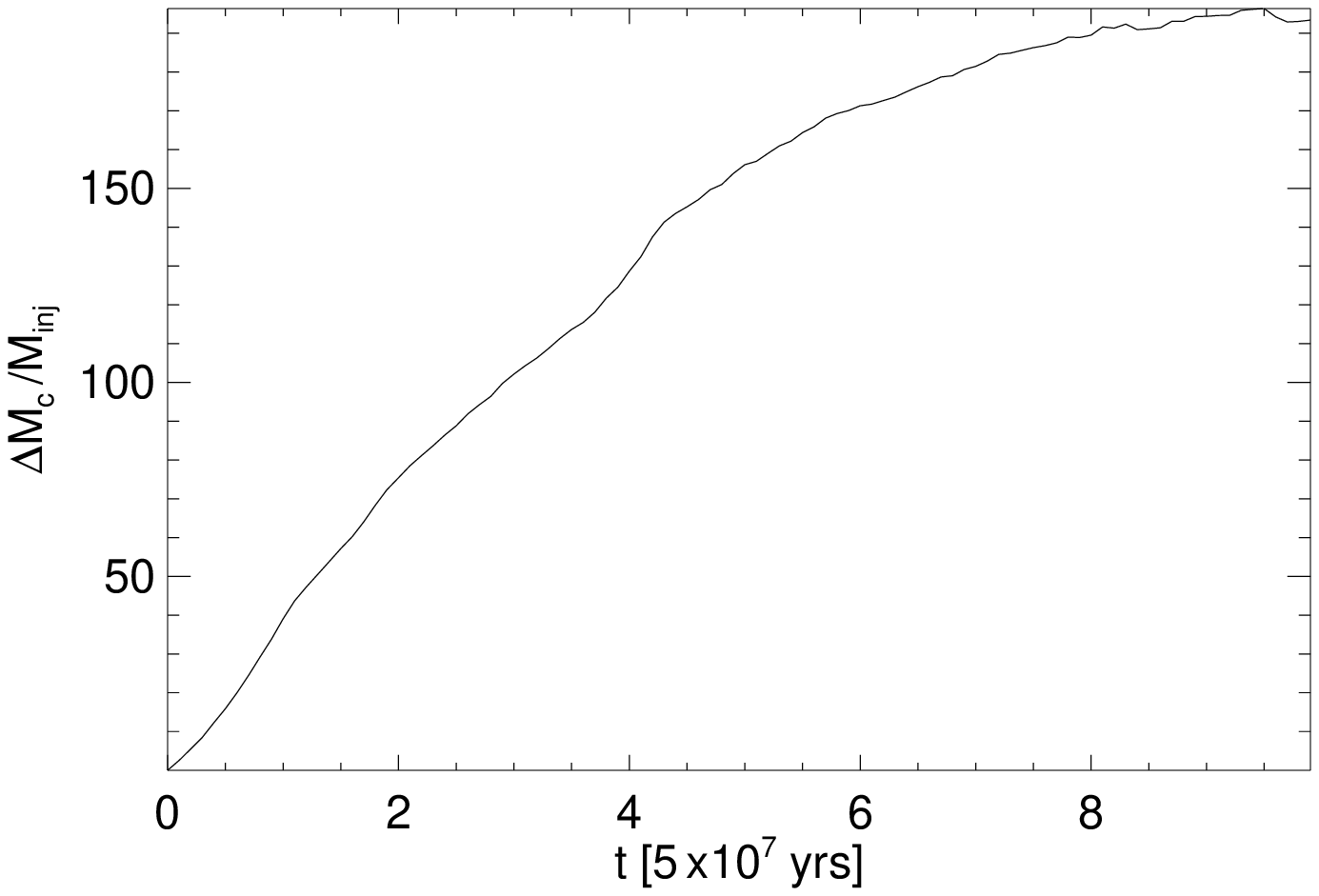}
}
\hbox{
\includegraphics[width=0.5\textwidth]{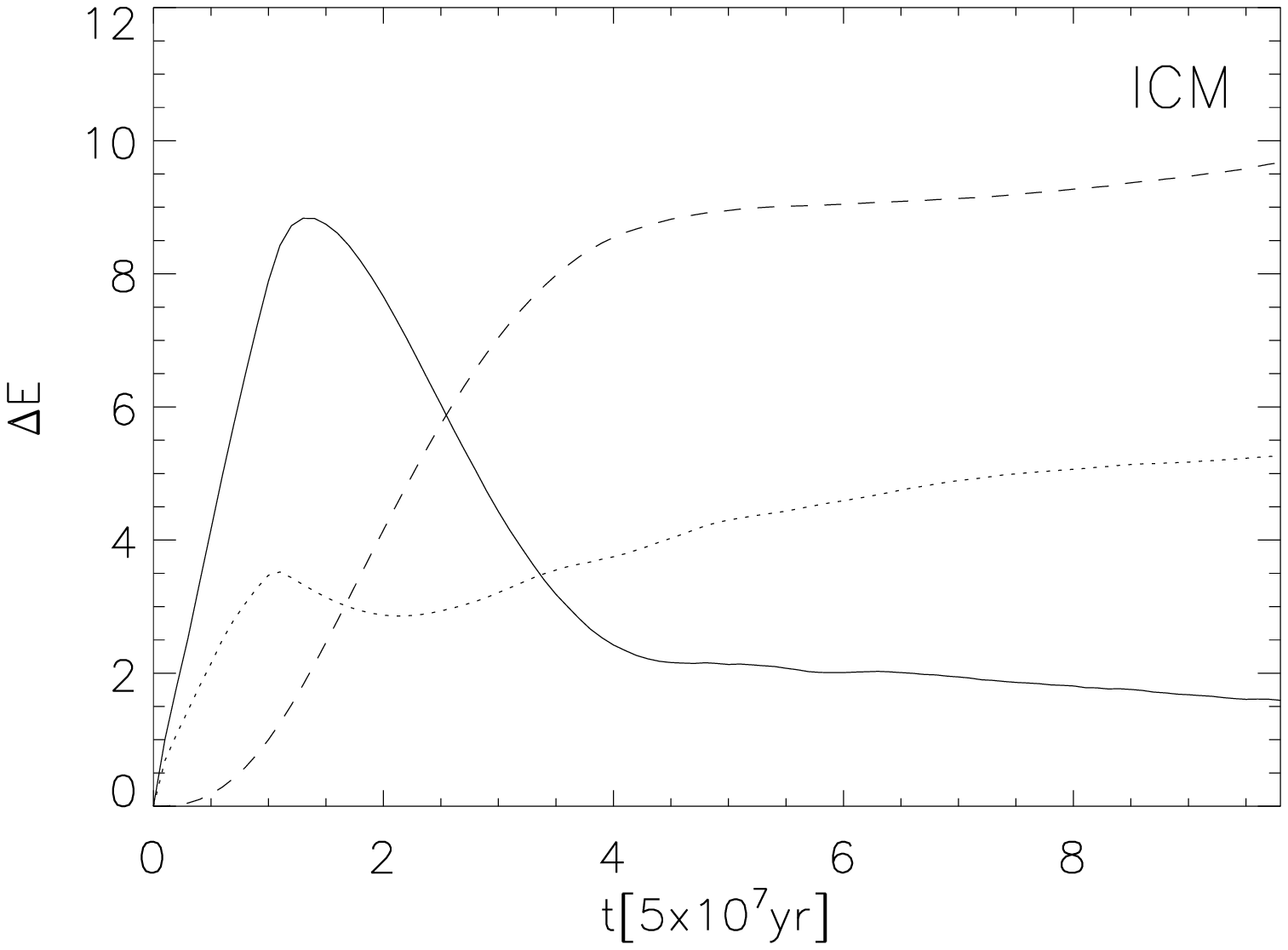}
\includegraphics[width=0.5\textwidth]{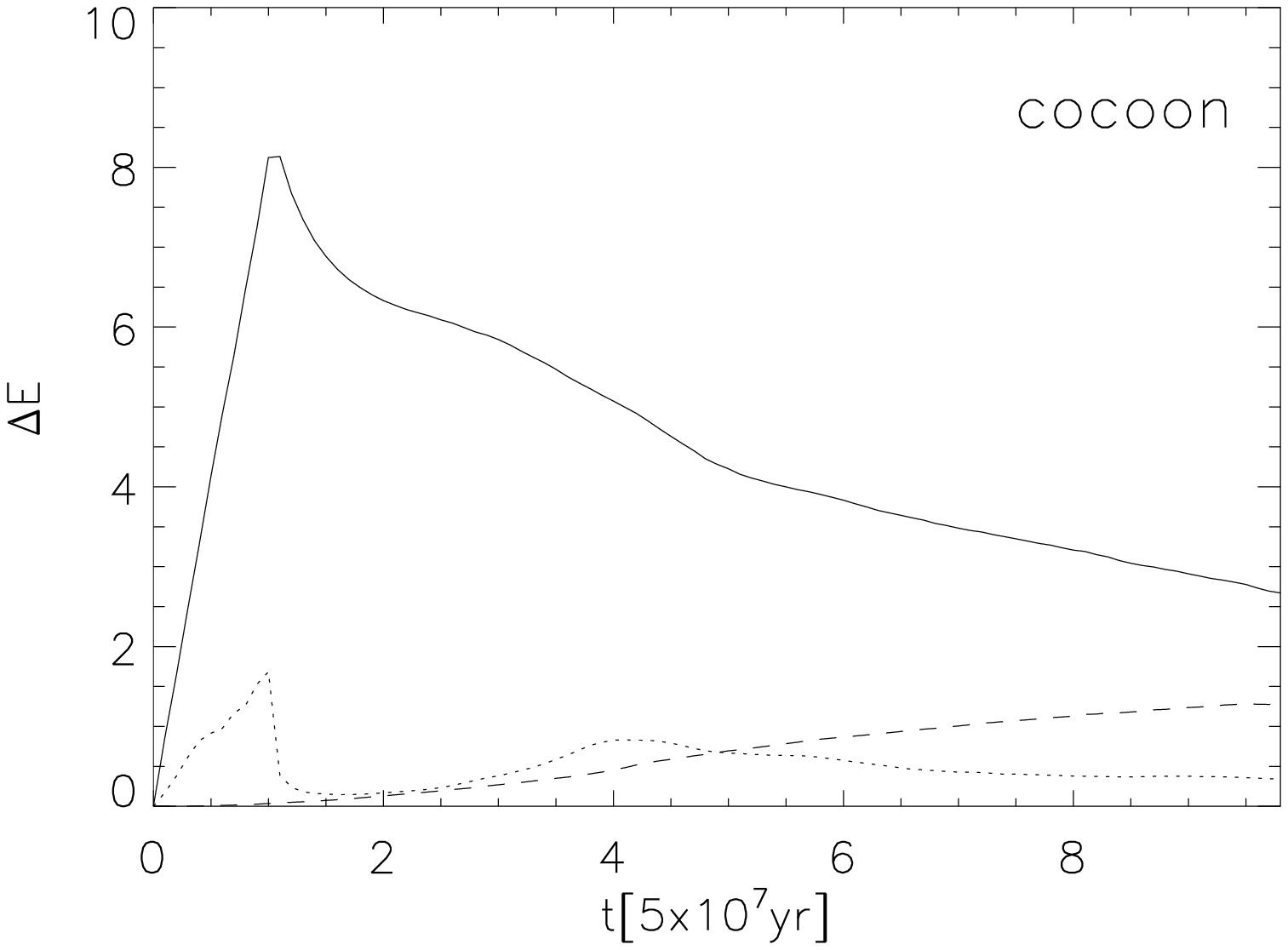}
}
\caption{Simulations diagnostics as a function of time.  {\it Top left
panel : }Volume averaged height of the plume as a function of time. {\it
Top right panel : }Ratio of mass entrained in the cocoon/plume against
mass injected by the radio jets.  {\it Bottom left panel : }Energetics
of the ambient ICM during the evolution of the radio galaxy.  Shown here
are the internal energy ($\Delta E_{\rm int,amb}$, solid line), kinetic
energy ($\Delta E_{\rm kin,amb}$, dotted line), and gravitational
potential energy ($\Delta E_{\rm pot,amb}$, dashed line).  {\it Bottom
right panel : }Energetics of the cocoon/plumes during the evolution of
the radio galaxy.  Shown here are the internal energy ($\Delta E_{\rm
int,coc}$, solid line), kinetic energy ($\Delta E_{\rm kin,coc}$, dotted
line), and gravitational potential energy ($\Delta E_{\rm pot,coc}$,
dashed line).
\label{fig:diagnostics}}
\end{figure*}

Continuing with the bubble analogy, we can write the buoyant velocity
$v_{\rm c}$ of the plume as
\begin{equation}
v_{\rm c}\approx {\cal C}\sqrt{\frac{4 r_{\rm bub} g (\rho_{\rm a}-\rho_{\rm
c})}{\rho_{\rm a}}}
\end{equation}
where $r_{\rm bub}$ the radius of the bubble, $g$ is the local
gravitational acceleration, and $\rho_{\rm a}$ and $\rho_{\rm c}$ are
the densities of the ambient and bubble materials respectively.  The top
left panel of Fig.~\ref{fig:diagnostics} shows the volume averaged plume
height (i.e., its distance from the center).  The plumes have
approximately constant velocity of $v_{\rm c}\sim 0.8$.  Together with
the cocoon size of $r_{\rm bub}(t = 4) \sim 1$ and the gravitational
acceleration at that position, this indicates that $\mathcal{C}$ is
slightly less than unity.

\subsection{Energetics and Thermodynamics}

One goal of these simulations is to determine a set of physical
parameters that are only weakly dependent on the details of the
simulations (most importantly, they should be weakly dependent on the
numerical resolution and the dimensionality).  In particular, we are
interested in the overall matter distribution, energetics, and
thermodynamics of the system.  For this discussion, it is interesting to
separate the system into different structural components, i.e., the
cocoon/plumes, and the ambient medium (including the shocked/compressed
shell that bounds the radio source).

Defining these regions is not trivial.  Since ZEUS-3D is an Eulerian
code we cannot, a priori, separate cocoon plasma from ambient plasma by
tracking fluid elements. Due to the turbulent cocoon surface, a simple
geometric identification of the cocoon is also not an option. However,
since the jet material is of very high entropy compared to the
background gas (especially once it has passed through the thermal
shock), we use the specific entropy $s=p/\rho^{5/3}$ as a discriminant.

\subsubsection{The entropy threshold method and numerical mixing}
\label{sec:diffusion}

For the simulation parameters we chose, the background material has
specific entropy (in code units)
\begin{equation}
s=\frac{p}{\rho^{5/3}}=\frac{3}{5}[1+(r/r_0)^2]^{1/2}.
\end{equation}
The injected material, on the other hand, has entropy $p/\rho^{5/3}
\sim 1300$ and cocoon material, which has gone through the terminal
shock, can have entropy as high as $s \sim 10^{5}$. We chose to define
the cocoon as the volume inside of which the entropy satisfies $s_{\rm
c} > 10$.  At all but the very earliest times in our simulations, the
shock driven into the ambient material is too weak to raise its
entropy above this threshold.  Conversely, an examination of entropy
maps from the simulations shows that, until about $t\sim 8$, the
contact discontinuity between the cocoon and ambient material is sharp
and well tracked by the $s_{\rm c}=10$ contour.  In fact, until $t\sim
5$, the location of the contact discontinuity is insensitive to the
chosen entropy threshold, provided that threshold is in the range
$10<s_{\rm c}< 100$.  For times $t>8$ numerical
mixing reduces substantial parts of the cocoon/plumes below the
$s_{\rm c}=10$ threshold.  At these late times, identification of a
well-defined contact discontinuity separating the plume from the
ambient material becomes problematic.

\begin{figure}
    \includegraphics[width=\columnwidth]{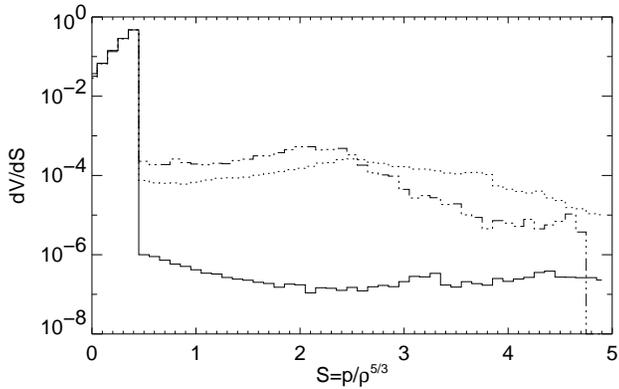}
    \caption{Normalized volume distribution of material with given specific
    entropies (computed over the whole computational domain).  The solid
    line corresponds to very early times $t=0.02$, the dotted line
    corresponds to $t=2$ and the dash-dotted line corresponds to $t=4$.
    The second broad peak in the distribution, which corresponds to the
    cocoon material, gradually moves to lower entropies as it is polluted
    by low entropy ISM/ICM gas.  \label{fig:s_distribution}}
\end{figure}

To better illustrate the effects of numerical mixing at later times, we
calculate the entropy distribution of Run~4.
Figure~\ref{fig:s_distribution} shows this function at different
times. Due to numerical mixing the second peak at high entropy (i.e.,
the cocoon material) diffuses to lower entropies.  It is clear that
after a certain amount of time an entropy cut is not a reasonable
estimator for cocoon material anymore, namely, when numerical diffusion
has lowered the entropy inside the cocoon to levels comparable with the
ambient entropy.

Numerical mixing is an irreversible process and hence creates entropy.  On
the other hand, `mixing' across the contact discontinuity in real radio
galaxies is not necessarily irreversible as can be seen from the following
argument.  Suppose that the ambient ISM/ICM material is not mixed on the
microscopic level.  Instead, assume that intact dense ISM/ICM filaments and
clouds are simply swept up into the light cocoon and kept distinct due to
being on different magnetic structures.  The fact that this is a
thermodynamically reversible process can be seen by noting that, given a
sufficiently long time, buoyancy effects could separate the ISM/ICM gas
from the cocoon plasma.  Thus, the entropy structure of the simulated
cocoon becomes increasingly less physical as time proceeds.  Indeed, once
numerical mixing has significantly affected a substantial part of the
simulated cocoon (which occurs at $t\sim 8$), the entropy structure of the
cocoon may be a poor reflection of reality.

\subsubsection{Definitions of energies}

To explore the energetics of our simulation in a quantitative manner,
we compute the following quantities:
\begin{enumerate}
\item Total mass of cocoon,
\begin{equation}
\hspace{2cm}M_{\rm cocoon}(t)=\int_{C}\rho {\rm d}V,
\end{equation}
\item Total internal energy of cocoon (ambient) material,
\begin{equation}
\hspace{2cm}E_{\rm int,coc(amb)}(t)=\frac{1}{\gamma-1}\int_{C(A)}p\,{\rm d}V,
\end{equation}
\item Total kinetic energy of cocoon (ambient) material
\begin{equation}
\hspace{2cm}E_{\rm kin,coc(amb)}(t)=\frac{1}{2}\int_{C(A)}\rho v^2\,{\rm d}V,
\end{equation}
\item Total gravitational potential energy of cocoon (ambient) material
\begin{equation}
\hspace{2cm}E_{\rm pot,coc(amb)}(t)=-\int_{C(A)}\rho\Phi\,{\rm d}V,
\end{equation}
\end{enumerate}
where $C$ is the region in which $s>10$, and $A$ is the region in which
$s\le 10$.  We then reference all energies to their values at the
initial time:
\begin{equation}
\Delta E=E(t)-E(0).
\end{equation}

\subsubsection{The active and early-inactive phases}

The bottom panels of Fig.~\ref{fig:diagnostics} shows how these
energies vary with time (upto $t=10$) for Run~4.  During the active
phase, the energy injected by the jets is transformed into internal
and kinetic energy of both the cocoon and ambient material, as well as
the potential energy of the ambient material.  All of these energy
forms increase linearly with time as the injected energy is shared
roughly according to equipartition.  After the source turns off, the
kinetic energy of the cocoon rapidly decreases as the rapid motions
associated with the jet cease.  During the short Sedov phase that
follows (between $t=1$ and $t\approx 1.5$), the internal energy of the
cocoon also undergoes a rapid decrease (adding mainly to the internal
energy of the ambient medium), as the overpressured cocoon expands
adiabatically to achieve pressure equilibrium with the ambient
material.  After the Sedov phase, the cocoon source then enters the
buoyant phase.

At all times during the buoyant phase in our simulations, the rising
buoyant plumes entrain and lift material (mainly in the trailing
regions).  This contributes to an increase in the gravitational
potential energy of both the plume and ambient material.
Concurrently, an approximately spherical pulse propagates radially
outwards in the ISM/ICM atmosphere.  This pulse is the remnant of the
strong shock that bounded the supersonic cocoon at early times.  As
will be discussed below, this pulse mediates a general expansion
(inflation) of the ISM/ICM atmosphere which results due to a heating
of the core regions by the jet activity.

\subsubsection{The entropy structure at late times}

\begin{figure*}
\hbox{
\includegraphics[width=0.5\textwidth]{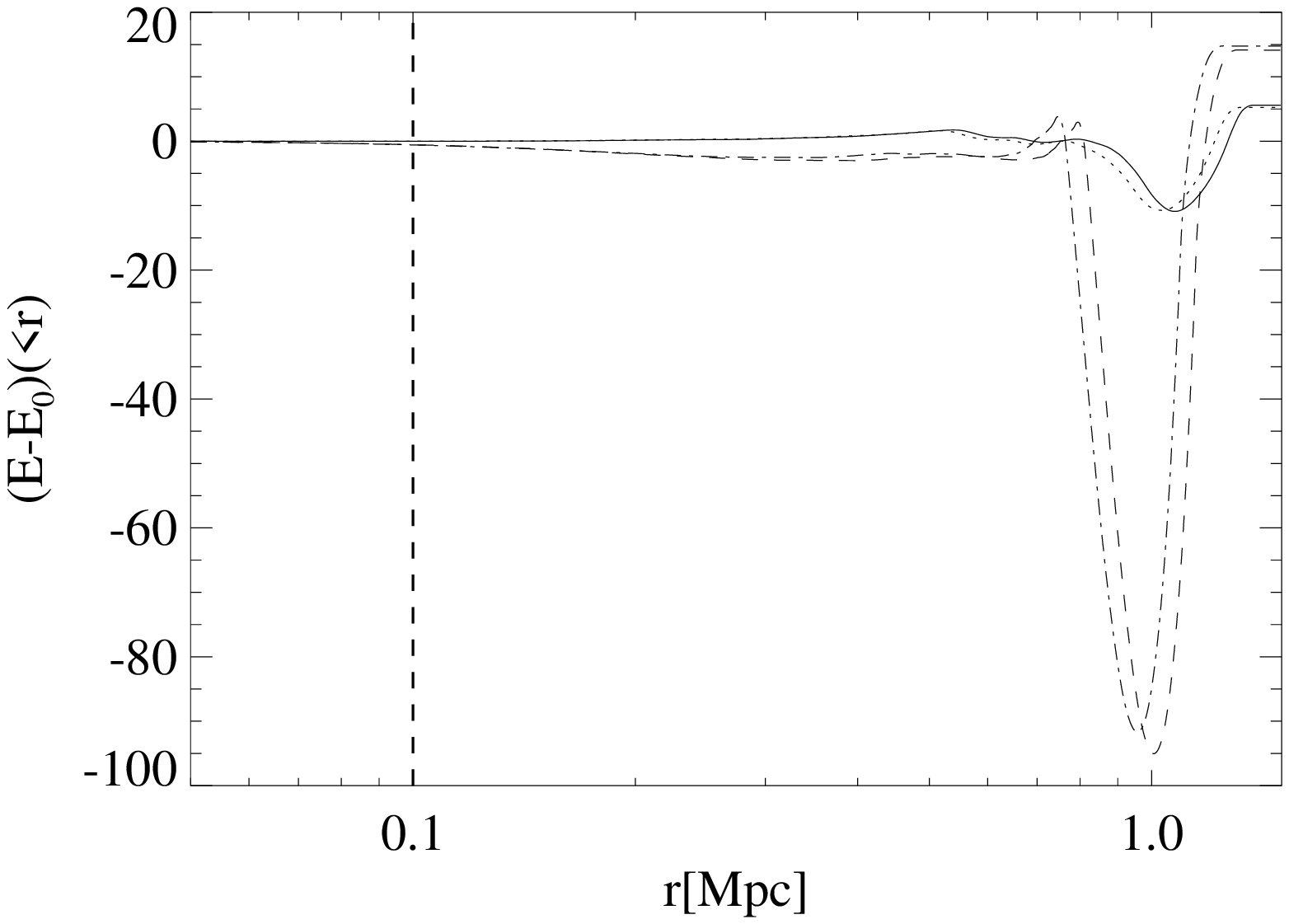}
\includegraphics[width=0.5\textwidth]{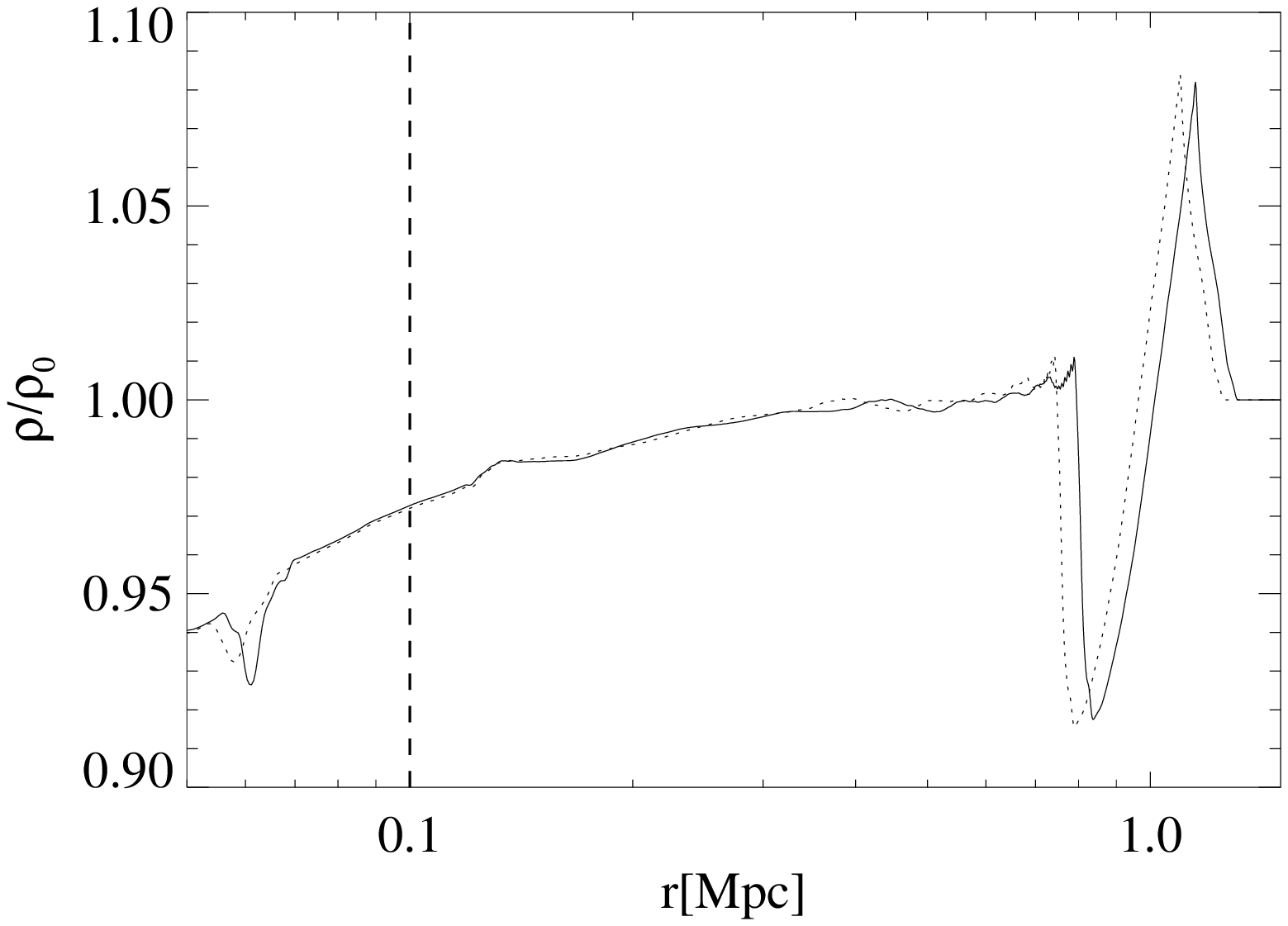}
}
\hbox{
\includegraphics[width=0.5\textwidth]{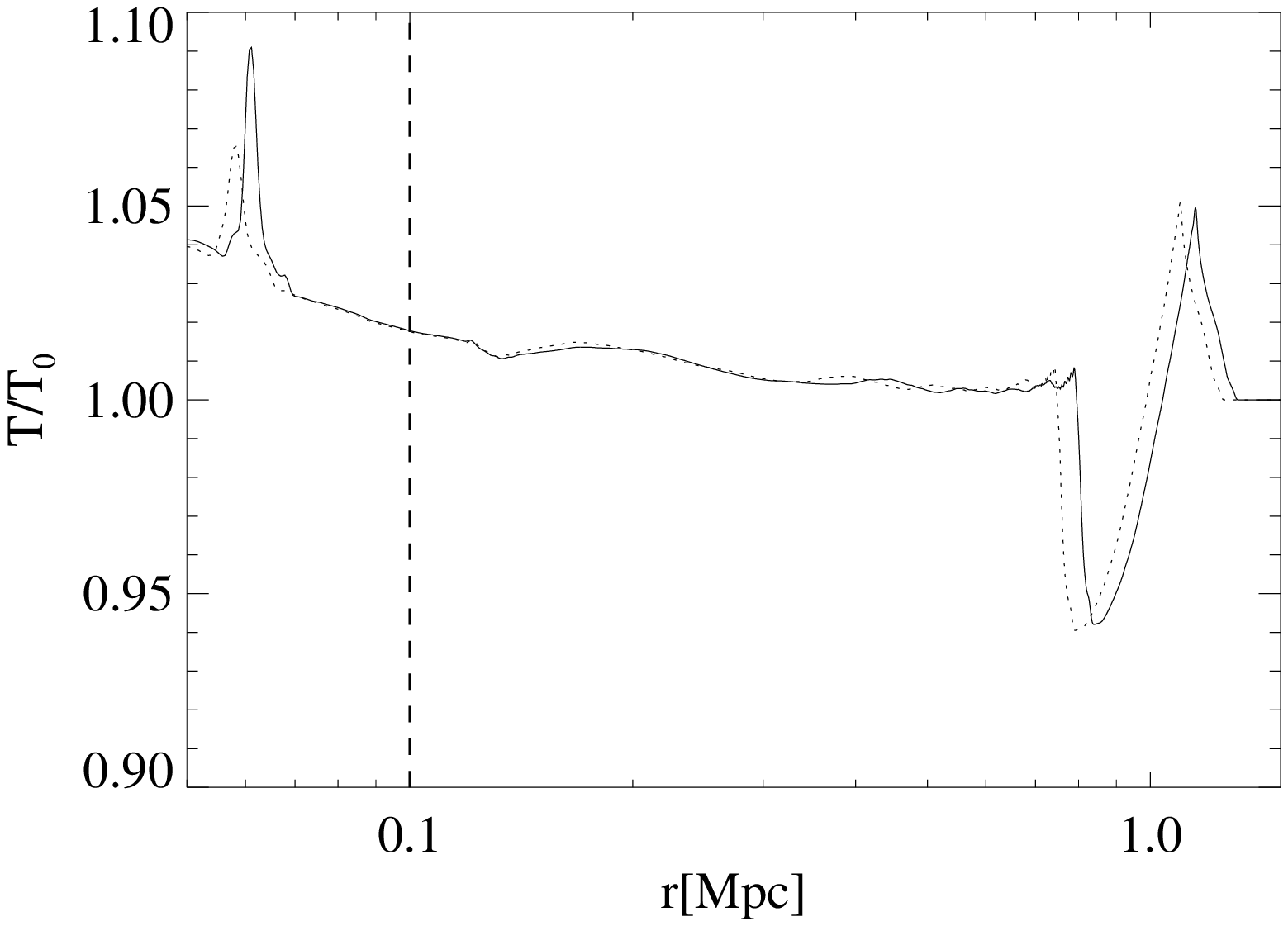}
\includegraphics[width=0.5\textwidth]{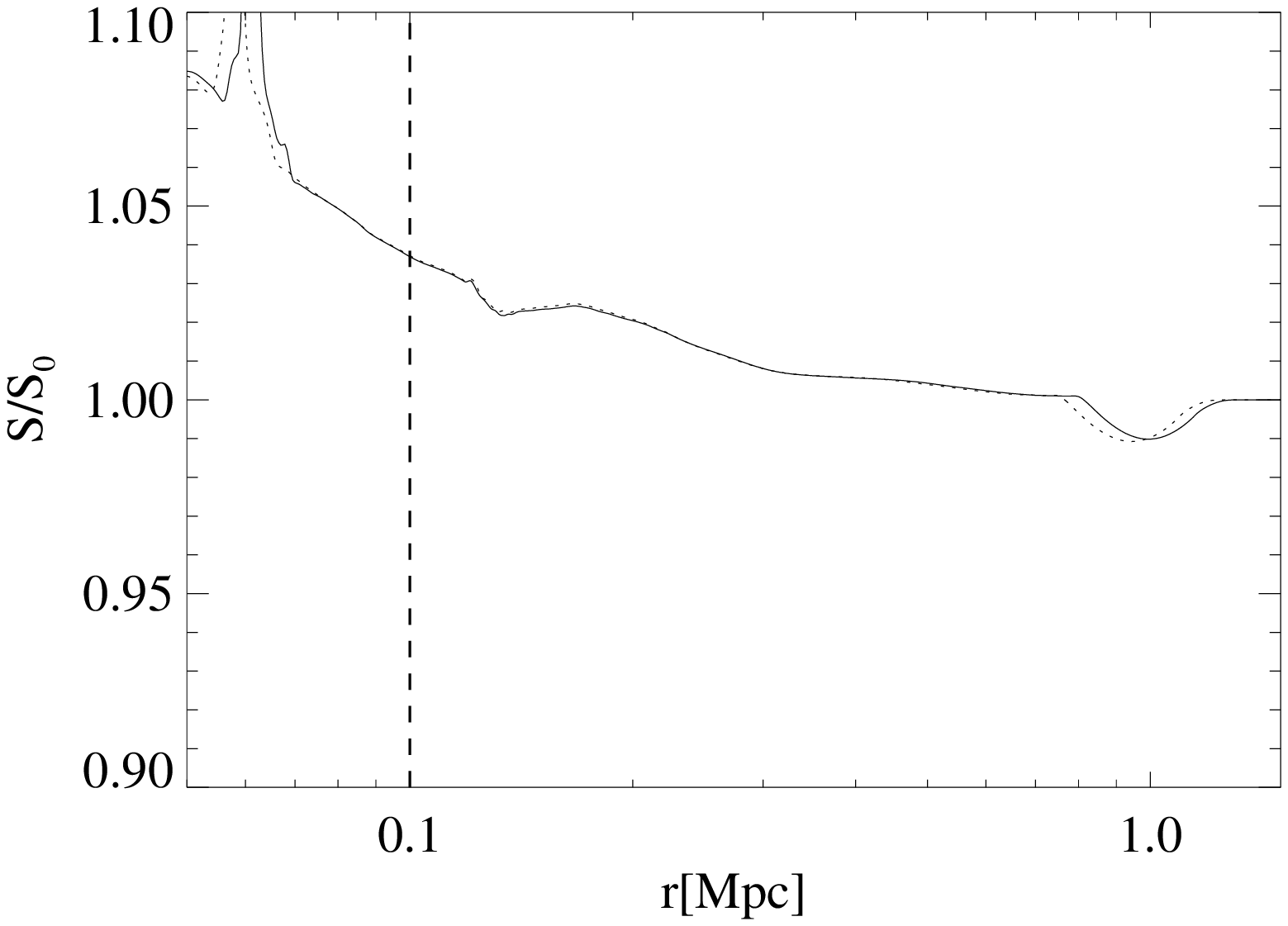}
}
\hbox{
\includegraphics[width=0.5\textwidth]{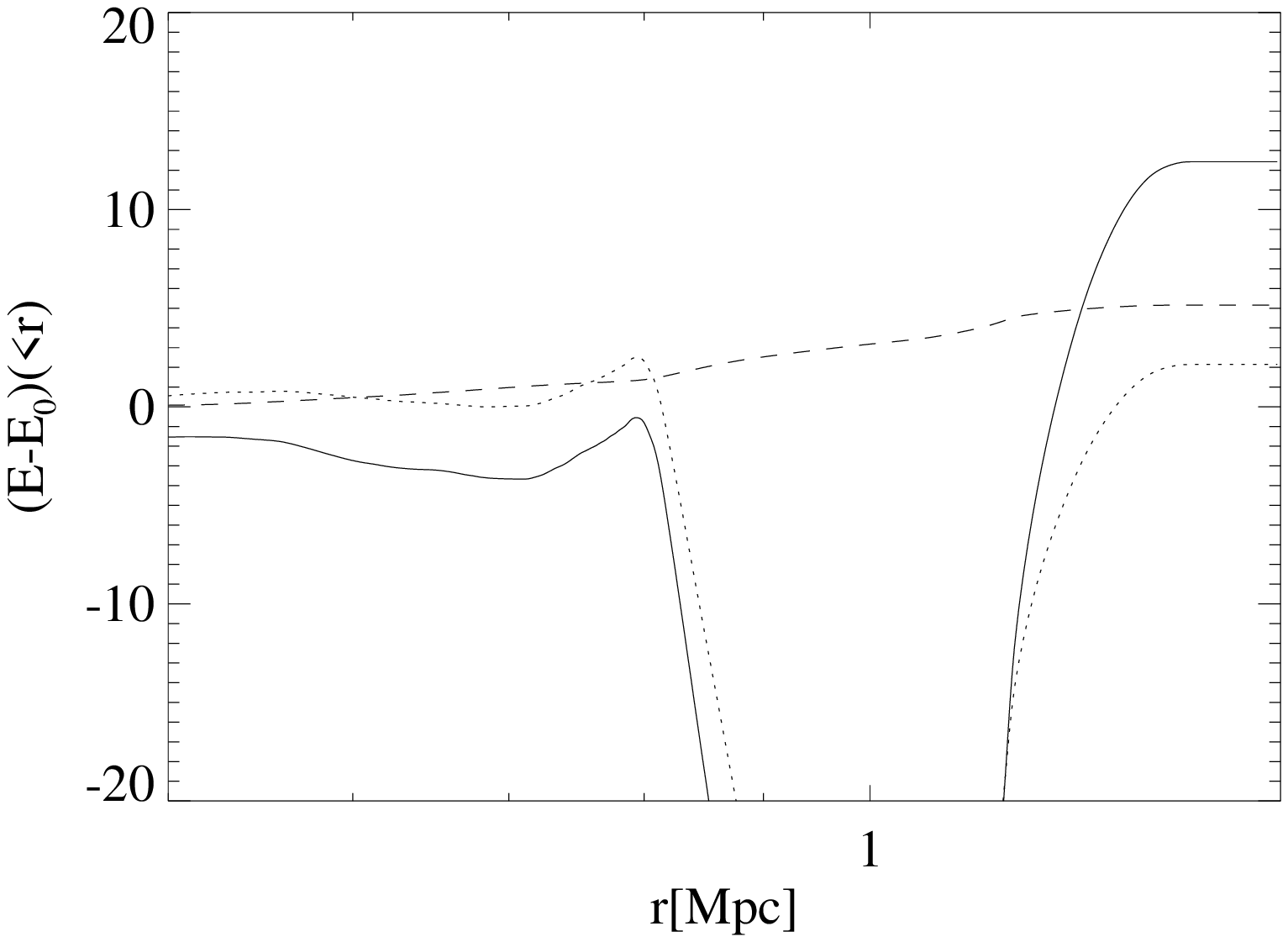}
\includegraphics[width=0.5\textwidth]{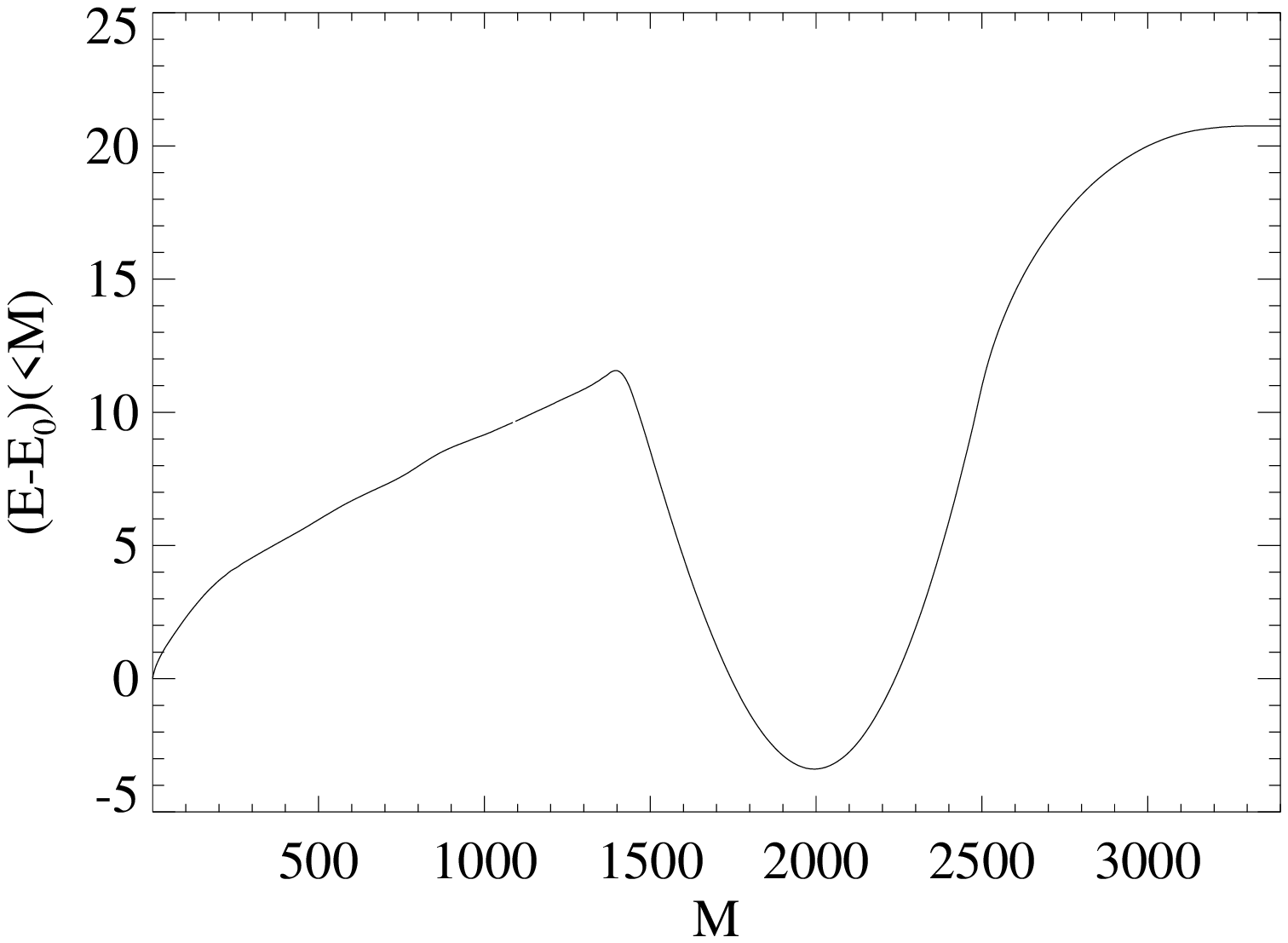}
}
\caption{A comparison of the initial and final states of the
simulation.  {\it Top left panel: }Total energy contained within
radius $r$ at two late times ($t=19$ and $t=20$) minus the same
quantity evaluated at the initial time ($t=0$).  The solid and dotted
lines show this energy-difference evaluated for times $t=20$ and
$t=19$, respectively, for regions along the jet axes
($\theta<30\degmark$ and $\theta>150\degmark$).  The dashed and
dot-dashed lines show this energy-difference evaluated for times
$t=20$ and $t=19$, respectively, for regions away from the jet axes
($30\degmark<\theta<150\degmark$).  The fact that these last curves
are higher at $r=30$ shows that most of the energy injected by the
radio galaxy resides in the expansion wave at large angles from the
jet axis.  {\it Top right panel : }Solid and dotted curves show the
angle-averaged density profile at $t=20$ and $t=19$, respectively,
ratioed against the initial density profile of the ambient material
for regions away from the jet axis.  {\it Middle left panel : }Solid
and dotted curves show the angle-averaged temperature profile at
$t=20$ and $t=19$, respectively, ratioed against the initial
(isothermal) temperature profile of the ambient material for regions
away from the jet axis.  {\it Middle right panel : }Solid and dotted
curves show the angle-averaged entropy profile at $t=20$ and $t=19$,
respectively, ratioed against the initial entropy profile of the
ambient material.  The entropy enhancement in the core of the ISM/ICM
atmosphere is clear.  The large fluctuations within the core radius
seen in these last three panels are due to the high-entropy blobs of
cocoon/plume material which are rising slowly under the action of
buoyancy forces.  The vertical dashed line in each plot denotes the
location of the core radius. {\it Bottom left panel : }Total energy
contained within radius $r$ at late time ($t=20$) minus the same
quantity evaluated at the initial time ($t=0$), with each of the three
forms of energy --- gravitational potential energy (solid line),
kinetic energy (dashed line) and internal energy (dotted line) ---
shown separately.  {\it Bottom right panel : }Total energy contained
within a given mass coordinate $m$ at late time ($t=20$) minus the
same quantity evaluated at the initial time ($t=0$).
\label{fig:profiles}}
\end{figure*}

The radio galaxy influences the entropy of the ISM/ICM atmosphere by
driving shocks into it.  At very long times after the jet activity has
ceased, the ICM/ISM will be in a relaxed hydrostatic equilibrium in
which the entropy is increasing outwards (or else there will be
convective turnover until such a state is achieved).  A parcel of gas
which passed through a very strong radio-galaxy induced shock (which
may have initially been in the centralmost regions of the
galaxy/cluster) will have a high final specific entropy and achieve
its new equilibrium state some large distance from the cluster core.
If the same parcel of gas suffers a weaker shock, the entropy increase
will be less and it will settle at a lower point in the final ISM/ICM
atmosphere.   

The density, temperature, and entropy plots in Fig. \ref{fig:profiles}
show that, after the episode of radio galaxy activity, the ISM/ICM
core is left on a higher adiabat than it was prior to the activity.
This is easy to understand.  Since the initial radio galaxy expansion
was supersonic, the material which initially had the lowest specific
entropy (i.e. at the center of the initial ISM/ICM distribution) has
all been shock heated.  The final ISM/ICM core contains a mixture of
(mildly) shocked gas, and unshocked ISM/ICM which has adiabatically
flowed inwards from more distant regions (and thus possesses a higher
specific entropy) to displace highly shocked material.  The highly
shocked material is expelled to the outer regions of the
galaxy/cluster.  For the specific case represented by our simulation,
the excess entropy over the undisturbed cluster gas peaks at 20\%.
This translates into an excess entropy of about 14\,${\rm
keV\,cm^{2}}$.

Although this is a small change in the entropy of the cluster core, one
might think that repeated radio galaxy activity would lead to a significant
accumulation of excess entropy in the cluster core.  This depends upon the
assumed entropy profile for the ISM/ICM.  Within the $\beta$-model
parameterization used here, the entropy profile of the ISM/ICM is rather
shallow.  Thus, even if one could imagine completely shocking a large part
of the central ISM/ICM, the entropy increase of the core seen in the final
equilibrium would still be rather modest since material just outside of the
strongly shocked region will simply flow inwards and take its place.  If
the initial entropy profile is much steeper (as it is in some strong
cooling flow models) then the same radio galaxy could impact the entropy of
the core rather more significantly.  Clearly, further work is needed on the
contribution of radio galaxies to the ISM/ICM entropy in which the entropy
injection is considered in relation to particular detailed cooling flow
models.

Using {\it ROSAT} data, Ponman, Cannon \& Navarro (1999) and Lloyd-Davies,
Ponman \& Cannon (2000) have found evidence for an ``entropy floor'' in
groups and clusters.  This suggests from an observational point of view
that these systems possess excess entropy (as compared to the entropy
profile expected from simple scaling relations and detailed simulations;
however, see alternative viewpoint of Bryan 2000).  Lloyd-Davies, Ponman
\& Cannon (2000) claim an excess entropy of 70--140\,${\rm keV\,cm^{2}}$
in a sample of 20 clusters and groups.  Interestingly, these authors also
conclude that this entropy injection occurred at low densities ($1-3\times
10^{-4}\pcmcu$) and hence they suggest that it occurred prior to the
collapse/formation of the cluster.  In principle, microscopic mixing of the
ICM and radio plasma {\it inside} of the cocoon/plumes might also lead to
ICM entropy injection at low density.  However, our simulations suggest
that buoyancy effects tend to transport any such mixed material completely
out of the cluster center.  Thus, if the determination of the density at
which the entropy injection occurred is robust, it does indeed seem likely
that the bulk of the entropy injection occurred prior to cluster formation
and is not due to more recent radio galaxy activity.  

\subsubsection{Energetics at late times}

The energetics of the very late stage of Run~4 are reported in
Fig.~\ref{fig:profiles}.  The large scale pulse mentioned in
\S4.2.3 manifests itself in each of the quantities plotted in
Fig.~\ref{fig:profiles} at $r\approx 20$.  The top-left panel shows
the difference in enclosed energies as a function of radius between
late times and the initial state $t=0$.  Comparing the values of this
function on each side of the pulse may suggest to the casual
reader that the wave contains most of the injected energy.  This is a
curious result since we have already noted appreciable entropy and
internal energy increases associated with the irreversible heating of
the ISM/ICM atmosphere.

The bottom panels of Fig.~\ref{fig:profiles} elucidate the true nature
of this pulse.  In the bottom-left panel, we note that the major
contributor to the increase of enclosed energy comes from
gravitational potential energy (rather than kinetic and internal
energy as would be expected if this were a sound wave carrying away
the injected energy).  Secondly, displaying the enclosed energy
difference in {\it mass coordinates} (Fig.~\ref{fig:profiles}
bottom-right panel) shows that the mass contained within the inner
edge of the wave-packet has acquired an appreciable excess of energy
during the evolution corresponding to at least half of the injected
energy.  

Thus, it can be seen that the ISM/ICM atmosphere has been inflated (or
puffed up) due to the heat deposited in the core of the atmosphere by
the jet activity.  The spherical pulse (travelling at the sound speed)
drives mass to a higher location in the potential thereby mediating
this inflation.  That this inflation has occurred is obvious from the
reduced core density in the final state (Fig.~\ref{fig:profiles} top
right).  Examining the enclosed energy in mass coordinates lets one
see that this inflation actually accounts for a substantial fraction
of the injected energy, implying that the injected energy has been
heated the ISM/ICM with a high efficiency ($\eta\sim 0.5$).

\section{Discussion}

\subsection{Cooling flows and radio galaxy heating}

As discussed above, we find that a large fraction ($\eta\sim 0.5$) of the
energy injected by the jet activity is thermalized in the core regions
of the ISM/ICM.  As mentioned in the Introduction, this raises the
possibility that radio galaxy activity can have significant effects on
the energetics of the ISM/ICM and, therefore, act as negative feedback
for cooling flow activity.

In particular, let us address the case of cluster cooling flows.
Peres et al. (1998) compared the radio and X-ray luminosity of 58
clusters in a flux limited {\it ROSAT} sample.  They found strong
radio sources in about $10\%$ of the clusters within the sample.
However, using the results of Peres et al. (1998) and summing over all
members of the sample (not just those with radio sources), we find
that the radio luminosity of the sample (defined as the sum over all
sources of $\nu L_\nu$ measured at 1.4\,GHz) is about 1\% of the
sample bolometric X-ray luminosity from within the ``cooling radius''.
To discuss the relevance of radio galaxy heating on the cluster, we
must relate the radio luminosity to the kinetic luminosity of the
sources.  

Simple dynamical models of expanding radio lobes suggest that the
maximum synchrotron luminosity of homogeneous, space-filling, lobes is
only a few percent of the kinetic luminosity (e.g., see Bicknell,
Dopita \& O'Dea 1997).  Departures from equipartition always act to
decrease the synchrotron efficiency, and the efficiency further
decreases as the source expands (Begelman 1996, 1999).  The only way
to increase this efficiency is to concentrate the radio emission into
a small volume of the radio lobe (e.g., strong shocks within the
lobe).  In the absence of such inhomogenieties, it seems likely that
the radio luminosity is at most a few percent of the kinetic
luminosity.  Therefore, using the Peres et al. (1998) results quoted
above, the sample-averaged kinetic luminosity is comparable to the
X-ray luminosity from within the cooling radius.  The results of our
work suggest that a large fraction of this jet power can be
thermalized in the ICM.

Thus, from a purely energetic point of view, radio galaxies can
balance the radiative cooling associated with cooling flows.  However,
it is clear that the ICM of real galaxy clusters does not experience a
steady-state situation in which jet-induced heating balances radiative
cooling.  As has already been noted, most of the integrated radio
luminosity in the Peres et al. (1998) sample is contributed by the
powerful radio galaxies that reside in only 10\% of the clusters.  The
arguments of the previous paragraph would suggest that, in these
clusters, the radio galaxy heating would vastly exceed the radiative
cooling.  Conversely, radiative cooling would certainly seem to
dominate in the remaining 90\% of the systems (unless they possess
powerful, but invisible, nuclear outflows; Binney 1999).  Thus, it is
clear that even if radio galaxy heating is relevant, the
heating/cooling balance can be vastly different from cluster to
cluster, or at different points of a given cluster's lifetime.

It is beyond the scope of this paper to confront, in any direct
manner, the plethora of new data that the {\it Chandra X-ray
Observatory} and {\it XMM-Newton} are providing on the properties of
the ICM and cooling flows.  Here, we simply note some of the
challenges that future models of cooling-flows and embedded
radio-galaxies must meet.  

{\it Chandra} finds spectroscopic evidence for cooling flows (albeit
with a mass deposition rate smaller than previously estimated with
{\it ROSAT}; Peres et al. 1998) even in clusters that contain
energetic radio galaxies.  For example, Fabian et al. (2000) find cool
gas in the Perseus cluster, even though the central radio galaxy is
clearly interacting with the ICM is a rather vigorous manner.
Curiously, the cool gas seems to form a shell around the radio lobes
and occupies precisely the location where we would expect
shock/compressionally heated gas to reside.  Furthermore, {\it
Chandra} has observed sharp temperature discontinuities in clusters
that do not appear to host powerful radio galaxies.  Firstly, a few
clusters (e.g., Abell 2142, Markevich et al. 2000) have large scale
temperature discontinuities which have become known as ``cooling
fronts''.  These may be discontinuities in the cooling flow associated
with a group/cluster merger event.  Secondly, Mazzotta et al. (2001)
has found a bubble of hot gas in the cluster MKW3S.  They associated
this with some previous outburst of nuclear activity from the central
galaxy.  However, if such a bubble had buoyantly risen from the centre
of the cluster, we would expect KH instabilities to disrupt it to a
larger extent than observed.  One would need to invoke some mechanism,
such as large scale magnetic fields, to stabilize the large scale KH
instabilities and prevent disruption of the bubble.

{\it XMM-Newton} has also produced a further mystery.  By performing
detailed emission line spectroscopy of several cooling flow clusters
(e.g., Abell 1835, Peterson et al. 2001; Abell 1795, Tamura et
al. 2001) with the reflection grating spectrometer (RGS) on {\it
XMM-Newton}, one can probe the detailed temperature distribution of
the ICM.  Multiple temperatures were observed in concordance with
cooling flow models, with components ranging from the virial
temperature down to $\sim 1$\,keV.  However, no emission lines were
observed from gas below $\sim 1$\,keV whereas cooling models would
predict large amounts of emission from such gas.  Fabian et al. (2001)
discusses various solutions for this puzzle and favours a model in
which there are large abundance inhomogenieties affecting the
predicting cooling rates.  Another solution is to postulate a heating
mechanism which targets cool ($\sim 1$\,keV) gas.  Examples of such a
mechanism are the ``percolation'' model of Begelman (2001), or the
``reconnection'' model of Norman \& Meiksin (1996).  Heating models
based upon simple hydrodynamic shocks (such as presented in this work
or that of David et al. 2001) do not possess such temperature
dependent targeting.  In order to match the observations, the putative
heating mechanism must meet the further challenge of heating the cool
gas rapidly up to the virial temperature.

Finally, we note that neither {\it Chandra} nor {\it XMM-Newton} have
yet found clear evidence for strong shocks associated with radio
galaxies.  Using the characterization of Paper I, all cluster radio
sources observed so far appear to be in the weak-shock or the no-shock
regime.  Although more work is required before making a rigorous
statement, there appears to be a disagreement between the observed
frequency of strong shocks and current hydrodynamic models of radio
galaxy evolution.  This might suggest that radio jets are less
powerful than we believe (i.e. the synchrotron efficiency is higher
than a few percent, implying that the radio emitting regions fill a
small portion of the radio lobes).  This problem was first pointed out
by David et al. (2001).

\subsection{The limitations of this work}

The simulations that we present in this paper are just the first step
in addressing the long-term evolution of radio galaxies.  In order to
make the problem tractable, we have made several simplifying
assumptions.  Here, we briefly discuss how these assumptions might
affect our results.

Our biggest simplification is that we have reduced the problem to two
spatial dimensions by assuming strict axisymmetry about the jet axis.
Given the inherent axisymmetry of the system, and the fact that
observed radio cocoons are approximately axisymmetric, this is not an
outrageous assumption.  However, there are several undesirable
consequences of assuming axisymmetry.  Firstly, we prevent the
formation of non-axisymmetric KH and RT modes.  All of the available
free energy is then channeled into the axisymmetric and RT modes.  The
axisymmetric KH modes are most efficient at mixing (since they present
the most effective barrier to the flow), and hence the axisymmetry
assumption maximizes the amount of mixing between the ambient and
cocoon material.  Secondly, the assumed axisymmetry prevents us from
including the effects of any realistic random ISM/ICM motions (i.e.,
``weather'').  These ICM motions will buffet the buoyantly rising
plumes and will become crucial in determining the morphology of very
old sources.

We have also chosen to ignore the presence of magnetic fields (i.e.,
we perform a hydrodynamic, rather than a magnetohydrodynamic,
calculation).  Magnetic fields in the radio cocoon might well be in
equipartition with the particle energies and hence could be
dynamically significant.  Dynamically important fields that are
tangled in an isotropic manner on small scales behave like a
relativistic gas and so can be handled within the context of a pure
hydrodynamic simulation.  However, anisotropic tangled fields, or
large scale magnetic fields, will introduce new features to the
dynamics which require a full magnetohydrodynamic treatment.
Similarly, the importance of neglecting the cocoon's relativistic
equation of state can only be assessed once calculations are performed
using relativistic hydrodynamic codes.

\section{Conclusions}

Radio galaxies are known to be dynamical systems that evolve on
timescales of $10^7-10^8\yr$.  Hence, we expect many galaxies and
clusters of galaxies to be host to dead, or relic, radio sources.  It
is possible, and maybe even probable, that this past radio activity
has influenced the energetics and thermodynamics of the hot,
space-filling medium in these hosts.  Thus, a study of the
environmental impact of dead radio galaxies is important and, given
the high quality X-ray data now coming from the {\it Chandra X-ray
Observatory} and {\it XMM--Newton}, very timely.

We have used high resolution hydrodynamic simulations to investigate one
particular scenario relevant to dying radio galaxies --- i.e., a rather
powerful radio-loud AGN situated at the center of a galaxy or cluster of
galaxies whose activity ceases abruptly.  In particular, we have simulated
back-to-back jets propagating in a $\beta$-model galaxy/cluster atmosphere.
We then shut down the jet activity and let the resulting structure evolve.
To make the problem more tractable, we assume axisymmetry about the jet
axis (thereby reducing the problem to two spatial dimensions) and neglect
the action of magnetic fields.  This is a significant extension of the work
of \C00 and \K00 since we simulate the active phase of the source and then
shut off the jets, rather than modelling the late stages of evolution by
simply letting a static bubble evolve under the action of buoyancy.

In the early lifetime of the simulated source (while the radio jets
are still active), the shocked jet material forms a cocoon which is
bounded by a shell of shocked ICM.  Even during the active phase,
hydrodynamic instabilities start to shred this cocoon.  However, only
after the jet activity ceases do KH and RT instabilities destroy the
integrity of the cocoon.  Thereafter, the old cocoon material forms
two plumes which rise in the cluster potential through the action of
buoyancy forces.  These plumes entrain a significant amount of cooler
material from the ICM core and lift this material high up into the
cluster atmosphere.  At very late times, the galaxy/cluster atmosphere
settles back into hydrostatic equilibrium but with a core specific
entropy that has been enhanced by $\sim 20\%$ over its initial value.
This entropy enhancement is due to the shocking of the lowest entropy
material by the strong shock which bounds the radio cocoon during the
early active phase.  We find that a large fraction ($\sim 0.5$) of the
injected energy is thermalized in the ISM/ICM.  Comparing late times
to the initial state, we find that the ISM/ICM atmosphere has
become inflated in order to maintain hydrostatic equilibrium
after the thermalization of the jets energy.  Such a large
thermalization efficiency raises the possibility that radio galaxies
are important in the overall energy budget of the ISM/ICM.

\section*{Acknowledgements}

CSR appreciates support from Hubble Fellowship grant HF-01113.01-98A.  This
grant was awarded by the Space Telescope Institute, which is operated by
the Association of Universities for Research in Astronomy, Inc., for NASA
under contract NAS 5-26555.  This work has also benefited from support by
NASA grant NAG5-7329.  Finally, we appreciate support from NASA under LTSA
grant NAG5-6337, and the National Science Foundation under grants
AST-9529170 and AST-9876887.

\begin{onecolumn}

\section*{Appendix}

In this appendix, we present a brief derivation of the relativistic
dispersion relation for the Kelvin-Helmholtz instability in
cylinderical geometry.  We then use this dispersion relation to
connect our simulations to the behavior of real sources.  Similar
relativistic KH analyses have been performed by many authors
(Blandford \& Pringle 1976; Ferrari, Trussoni \& Zaninetti 1978; Ray
1981; Birkinshaw 1984; Zaninetti 1985).  We repeat the derivation here
for completeness.

The continuity and momentum equations of relativistic fluid dynamics for a
$\gamma$-law gas are
\begin{eqnarray}
\frac{\partial}{\partial x^i}(\rho u^i)&=&0\\
hu^k\frac{\partial u_i}{\partial x^k}-\frac{\partial p}{\partial x^i} +
u_i u^k \frac{\partial p}{\partial x^k}&=&0\\
u^k\frac{\partial }{\partial x^k}(p\rho^{-\gamma})&=&0,
\end{eqnarray}
where $\rho$ and $p$ are the comoving density and pressure, respectively,
$u^i$ are the components of the four velocity, and $\gamma$ is the
polytropic index.  Here, $h=\rho+\gamma p$ is the relativistic enthalpy.
We wish to use these equations to study the relativistic Kelvin-Helmholtz
instability.

Consider first-order perturbations of these equations:
\begin{eqnarray}
p&=&p_0+\Delta p\\
\rho&=&\rho_0+\Delta\rho\\
u^i&=&u^i_0+\Delta u^i.
\end{eqnarray}
For simplicity, we shall drop the subscript `0' in the subsequent
equations.  Assume that the unperturbed quantities are time-independent and
spatially uniform.  The linearized fluid equations can then be cast in the
form
\begin{equation}
\left(\frac{h}{\gamma p}-1\right)u^i u^k \frac{\partial^2 \Delta p}{\partial
x^k\partial x^i} = -\frac{\partial^2 \Delta p}{\partial
x^i\partial x_i}
\end{equation}
For definiteness, we choose cylinderical polar coordinates and say that the
unperturbed velocity field is along the $z$-direction with velocity $V$.
This last equation can then be written as
\begin{equation}
\Gamma^2\left(\frac{h}{\gamma
p}-1\right)\left(\frac{\partial^2}{\partial t^2}
+2V\frac{\partial^2}{\partial t \partial z}+
V^2\frac{\partial^2}{\partial z^2}\right)\Delta p = \left(\nabla^2
-\frac{\partial^2}{\partial t^2} \right)\Delta p.
\end{equation}

Now consider the following geometry which is appropriate for the analysis
of the Kelvin-Helmholtz instability in radio galaxy cocoons.  Consider a
contact discontinuity situated at the surface $r=R$.  For $r<R$, we have a
fluid (denoted by subscript 1) which has an unperturbed velocity $V_1$ in
the $z$-direction.  For $r>R$, we have a fluid (denoted by subscript 2)
which has an unperturbed velocity of $V_2$.  We search for perturbed modes
of the form $\Delta p=f(r)\exp[i(kz+m\phi-\omega t)]$, imposing a
regularity condition at $r=0$ and only permitting outgoing waves in the
region $r>R$.  Matching the pressure perturbations at $r=R$ gives the
following dispersion relation:
\begin{equation}
\frac{J_m(K_1 R)}{K_1J^\prime_m (K_1R)}\Gamma_1h_1(\omega-kV_1)^2=
\frac{H^{(1)}_m(K_2 R)}{K_2H^{(1)\prime}_m (K_2R)}\Gamma_2h_2(\omega-kV_2)^2,
\end{equation}
where
\begin{equation}
K_1^2=\Gamma_1^2\left(\frac{h_1}{\gamma_1
p}-1\right)(\omega-kV_1)^2-k^2+\omega^2,
\end{equation}
with a similar expression for $K_2$.  Without loss of generality, we can
suppose that the fluid in the region $r>R$ is at rest ($V_2=0$,
$V_1=V$).  It can be seen from this last expression with a little
algebra that, in limit of small wavelengths and only mildly relativistic
wave speeds, the general solution has the form
\begin{equation}
\omega=kf\left(h_1/h_2,{\cal M},\gamma_1,\gamma_2\right)
\end{equation}
where ${\cal M}=V/c_{\rm s,2}$, where $c_{\rm s,1}$ is the sound speed
in the region $r<R$.

We can now use this general solution to comment on the connection
between our simulations and the hydrodynamics of real sources.  Since
most of the interesting hydrodynamics is associated with the development
of the KH instability, we should endeavour to approximate the expected
real-life KH growth rates within our simulations.

The cocoons in real radio galaxies will, at least initially, be
dominated by plasma with a relativistic equation of state
(i.e. $\gamma_1=4/3$).  On the other hand, our simulations are
inherently non-relativistic (i.e., $\gamma_1=5/3$).  While fully
relativistic simulations are needed to address, in full, the effect of a
relativistic equation of state, we can use the above dispersion relation
to judge the effect on the KH growth rates of neglecting the
relativistic nature of the cocoon plasma.  Numerical solutions of eqn
(19) show that the KH growth rates and wavelengths are little affected
by changing from $\gamma_1=4/3$ to $\gamma_1=5/3$.  Hence, neglecting
the relativistic equation of state is not a serious flaw of our
simulations.

Noting that the Mach number of the cocoon backflow is determined by the
hydrodynamics of the system (and is always approximately unity), the
remaining parameter relevant to the KH growth rate is the ratio
$h_1/h_2$.  In real systems, the ambient thermal material will be
non-relativistic (with sound speed $c_{\rm s,2}$) while the cocoon
(especially near the jet head) will be dominated by relativistic plasma
($h_1\approx p$).  Since $c\gg c_{s,2}$, the ratio of relativistic
enthalpies will be
\begin{eqnarray}
\frac{h_1}{h_2}&=&\frac{p}{\rho_2\,c^2}\\
&=&\left(\frac{c_{s,1}}{c}\right)^2.
\end{eqnarray}
Plausible parameters give $c_{s,1}\sim 3000-4000\kmps$, giving
$h_1/h_2\sim 1-2\times 10^{-4}$.  Any pollution of the relativistic
plasma in the cocoon (due to mixing) will reduce this ratio appreciably.
For comparison, in our simulations we have
\begin{eqnarray}
\frac{h_1}{h_2}&=&\frac{\rho_1}{\rho_2}\\
&=&\left(\frac{c_{\rm s,2}}{c_{\rm s,1}}\right)^2,
\end{eqnarray}
and by measuring the sound speeds, we determine that we achieve
$h_1/h_2\sim 5-10\times 10^{-4}$.  We conclude that while it would be
desirable to push this ratio lower by a factor of a few, our simulations
are exploring the appropriate parts of parameter space.

\end{onecolumn}


\begin{thebibliography}{}
\bibitem{}Arnaud K.~A., Fabian A.~C., Eales S.~A., Jones C., Forman W.,
1984, MNRAS, 211, 981
\bibitem{}Begelman M.~C., 2001, to appear in Particles and Fields in Radio 
Galaxies, ed. R.~A.~Laing \& K.~M.~Blundell.
\bibitem{}Begelman M.~C., 1999, in The Most Distant Radio Galaxies, ed.
H.~J.~A.~R\"ottering, P.~N.~Best \& M.~D.~Lehnert (Amssterdam: Royal 
Netherlands Academy of Arts and Sciences), 173 
\bibitem{}Begelman M.~C., 1996, in Cygnus~A -- Study of a Radio Galaxy, 
ed. C.~Carilli \& D.~Harris (Cambridge: Cambridge University Press), 209
\bibitem{}Begelman M.~C., Cioffi D.~F., 1989, ApJ, 345, L21
\bibitem{}Bicknell G.~V., Dopita M.~A., O'Dea C.~P., 1997, ApJ, 485, 112
\bibitem{}Binney J., 1999, in The Radio Galaxy M87, ed. H-.~J.R\"oser \& 
K.~Meisenheimer, Springer-Verlag, 116
\bibitem{}Birkinshaw M., 1984, MNRAS, 208, 887
\bibitem{}Blandford R.~D., Pringle J.~E., 1976, MNRAS, 176, 443
\bibitem{}Br\"uggen M., Kaiser C.~R., 2000, MNRAS, submitted
\bibitem{}Bryan G., 2000, ApJL, in press
\bibitem{}Cesarsky C.~J., Kulsrud R.~M., 1981, in Origin of Cosmic Rays,
IAU Symp. 94., ed. G.~Setti, G.Spada, A.~W., Wolfendale, (Dordrecht:
Reidel), 251
\bibitem{}Churazov E., Br\"uggen M., Kaiser C.~R., B\"ohringer H., Forman
W., 2000, submitted to ApJ (astro-ph/0008215).
\bibitem{}Cioffi D.~F., Blondin J.~M., 1992, ApJ, 392, 458
\bibitem{}Clarke D.~A., Norman M.~L., Fiedler R.~A., 1994, National
Center for Supercomputing Applications Technical Report 15.
\bibitem{}David L.~R. et al., 2001, ApJ, 557, 546
\bibitem{}Etorri S., Fabian A.~C., 2000, MNRAS, in press
\bibitem{}Fabian A.~C., Mushotzky R.~F., Nulsen P.~E.~J., Peterson J.~R., 
2001, MNRAS, 321, L20
\bibitem{}Fabian A.~C. et al., 2000, MNRAS, 318, L65
\bibitem{}Faranoff B.~L., Riley J.~M., 1974, MNRAS, 167, L31
\bibitem{}Ferrari A., Trussoni E., Zaninetti L., 1978, A\&A, 64, 43
\bibitem{}Gull S.~F., Northover K.~J.~E., 1973, Nature, 224, 80
\bibitem{}Hooda J.~S., Mangalam A.~V., Wiita P.~J., 1994, ApJ, 423, 116
\bibitem{}Lind K.~R., Payne D.~G., Meier D.~L., Blandford R.~D., 1989,
ApJ, 344, 89
\bibitem{}Lloyd-Davis E.~J., Ponman T.~J., Cannon D.~B., 2000, MNRAS,
in press
\bibitem{}Markevich M. et al., 2000, ApJ, 541, 542
\bibitem{}Mazzotta P., Kaastra J.~S., Paerels F.~B., Ferrigno C., Colafrancesco S., Mewe R., Forman W.~R., 2001, ApJL, in press
\bibitem{}Norman C., Meiksin A., 1996, ApJ, 468, 97
\bibitem{}Norman M.~L., Winkler K.~-H.~A., Smarr L., Smith M.~D., 1982,
A\&A, 113, 285
\bibitem{}Peres C.~B., Fabian A.~C., Edge A.~C., Allen S.~W., Johnstone R.~M., White D.~A., 1998, MNRAS, 298, 416
\bibitem{}Peterson J.~A. et al., 2001, A\&A, 365, 104
\bibitem{}Ponman T.~J., Cannon D.~B., Navarro J.~F., 1999, Nature, 397, 135
\bibitem{}Ray T.~P., 1981, MNRAS, 196, 195
\bibitem{}Reynolds C.~S., Fabian A.~C., 1996, MNRAS, 278, 479
\bibitem{}Reynolds C.~S., Heinz S., Begelman M.~C., 2001, ApJL, in press
\bibitem{}Scheuer P.~A.~G., 1974, MNRAS, 166, 513
\bibitem{}Stone J.~M., Norman M.~L., 1992a, ApJS, 80, 791
\bibitem{}Stone J.~M., Norman M.~L., 1992b, ApJS, 80, 81
\bibitem{}Tamura T. et al., 2001, A\&A, 365, L87
\bibitem{}Zaninetti L., 1985, A\&A, 1985, 79, 82
\end{thebibliography}
\end{document}